\documentclass[12pt,eqno,epsf]{article}
\usepackage{amsmath,amssymb,graphicx,slashbox}
\numberwithin{equation}{section}

\def\ignore#1{{}}

\newcounter{sxn}

\newcounter{axn}

\date{}

\newdimen\mybaselineskip
\renewcommand{\thefootnote}{\arabic{footnote}}

\newcommand{\beeq}{\begin{equation}}
\newcommand{\eneq}{\end{equation}}
\newcommand{\beqn}{\begin{eqnarray}}
\newcommand{\eeqn}{\end{eqnarray}}

\newcommand{\alp}{\alpha}
\newcommand{\bt}{\beta}
\newcommand{\gm}{\gamma}
\newcommand{\Gm}{\Gamma}
\newcommand{\dlt}{\delta}

\newcommand{\ep}{\epsilon}
\newcommand{\vep}{\varepsilon}
\newcommand{\tht}{\theta}
\newcommand{\vth}{\vartheta}

\newcommand{\lmd}{\lambda}
\newcommand{\Lmd}{\Lambda}
\newcommand{\sgm}{\sigma}

\newcommand{\vph}{\varphi}
\newcommand{\omg}{\omega}
\newcommand{\Omg}{\Omega}

\newcommand{\be}{\begin{equation}}
\newcommand{\ee}{\end{equation}}
\newcommand{\bea}{\begin{eqnarray}}
\newcommand{\eea}{\end{eqnarray}}
\newcommand{\eql}{\!\!\!&=\!\!\!&}

\newcommand{\sma}{\!\!\!&\simeq\!\!\!&}
\newcommand{\defa}{\!\!\!&\equiv\!\!\!&}

\newcommand{\toa}{\!\!\!&\to\!\!\!&}

\newcommand{\simlt}{\stackrel{<}{{}_\sim}}

\newcommand{\tl}[1]{\tilde{#1}}
\newcommand{\bdm}[1]{{\mbox{\boldmath $#1$}}}
\newcommand{\tr}{{\rm tr}}

\newcommand{\diag}{{\rm diag}}
\newcommand{\der}{\partial}
\newcommand{\dr}{\!\!d}

\newcommand{\ie}{{\it i.e.}}

\newcommand{\brkt}[1]{\left( #1 \right)}
\newcommand{\brc}[1]{\left\{ #1 \right\}}
\newcommand{\sbk}[1]{\left[ #1 \right]}
\newcommand{\abs}[1]{\left| #1 \right|}

\newcommand{\ket}[1]{|#1\rangle}

\renewcommand{\Re}{{\rm Re}\,}
\renewcommand{\Im}{{\rm Im}\,}

\newcommand{\cA}{{\cal A}}
\newcommand{\cB}{{\cal B}}

\newcommand{\cD}{{\cal D}}

\newcommand{\cL}{{\cal L}}
\newcommand{\cM}{{\cal M}}

\newcommand{\cO}{{\cal O}}
\newcommand{\cP}{{\cal P}}

\newcommand{\cS}{{\cal S}}
\newcommand{\cT}{{\cal T}}

\newcommand{\cW}{{\cal W}}

\newcommand{\ubl}{U(1)_X}
\newcommand{\suL}{SU(2)_{\rm L}}
\newcommand{\suR}{SU(2)_{\rm R}}
\newcommand{\uy}{U(1)_Y}
\newcommand{\uem}{U(1)_{\rm EM}}
\newcommand{\mKK}{m_{\rm KK}}
\newcommand{\aL}{a_{\rm L}}
\newcommand{\aR}{a_{\rm R}}
\newcommand{\ha}{\hat{a}}

\newcommand{\thH}{\theta_{\rm H}}

\newcommand{\NP}[1]{{\it Nucl.~Phys.}~{\bf #1}}
\newcommand{\PL}[1]{{\it Phys.~Lett.}~{\bf #1}}

\newcommand{\MPL}[1]{{\it Mod.~Phys.~Lett.}~{\bf #1}}

\newcommand{\PR}[1]{{\it Phys.~Rev.}~{\bf #1}}
\newcommand{\PRL}[1]{{\it Phys.~Rev.~Lett.}~{\bf #1}}
\newcommand{\PTP}[1]{{\it Prog.~Theor.~Phys.}~{\bf #1}}

\newcommand{\JH}[1]{{\it JHEP}~{\bf #1}}

\begin{document}
\thispagestyle{empty}

\baselineskip=12pt

{\small \noindent 
\hfill OU-HET 634/2009}

{\small \noindent \hfill  RIKEN-TH-163}

\baselineskip=35pt plus 1pt minus 1pt

\vskip 2.0cm

\begin{center}
{\Large \bf Tree-level unitarity in Gauge-Higgs Unification}\\

\vspace{2.0cm}
\baselineskip=20pt plus 1pt minus 1pt

\normalsize

{\bf Naoyuki\ Haba},${}^1\!${\def\thefootnote{\fnsymbol{footnote}}
\footnote[1]{\tt e-mail address: haba@het.phys.sci.osaka-u.ac.jp}} 
{\bf Yutaka\ Sakamura}${}^2\!${\def\thefootnote{\fnsymbol{footnote}}
\footnote[2]{\tt e-mail address: sakamura@riken.jp}}
{\bf and
Toshifumi\ Yamashita}${}^3\!${\def\thefootnote{\fnsymbol{footnote}}
\footnote[3]{\tt e-mail address: yamasita@eken.phys.nagoya-u.ac.jp}}

\vspace{.3cm}
${}^1${\small \it Department of Physics, Osaka University, 
Toyonaka, Osaka 560-0043, Japan} \\
${}^2${\small \it RIKEN, Wako, Saitama 351-0198, Japan} \\
${}^3${\small \it Department of Physics, Nagoya University, 
Nagoya 464-8602, Japan}
\end{center}

\vskip 2.0cm
\baselineskip=20pt plus 1pt minus 1pt

\begin{abstract}
We numerically estimate a scale~$\Lmd_{\rm uni}$ 
at which tree-level unitarity is violated 
in the $SO(5)\times\ubl$ gauge-Higgs unification model 
by evaluating amplitudes for scattering of the longitudinal W bosons. 
The scattering amplitudes take larger values in the warped spacetime 
than in the flat spacetime, and take maximal values when $\thH=\pi/2$, 
where $\thH$ is the Wilson line phase along the extra dimension. 
We take into account not only the elastic scattering but also 
possible inelastic scatterings in order to estimate $\Lmd_{\rm uni}$. 
We found that $\Lmd_{\rm uni}\simeq 1.3\mKK$ 
in the warped spacetime, and $\Lmd_{\rm uni}\simeq 140\mKK$ 
in the flat spacetime, where $\mKK$ is the Kaluza-Klein mass scale.  
The tree-level unitarity is violated at $\cO(1~\mbox{TeV})$ for $\thH=\pi/2$ 
in the former case due to the vanishing $WWH$ coupling. 
\end{abstract}


\newpage

\section{Introduction}
Extra dimensions are interesting candidates of the physics 
beyond the standard model (SM), and have been extensively investigated 
during the past decade. 
They open up new possibilities for various issues, such as 
the large hierarchy between the electroweak and Planck scales~\cite{ADD,RS} 
or among the fermion masses~\cite{ArkaniSchmaltz}, 
a mechanism of gauge symmetry breaking~\cite{Kawamura}, 
candidates of dark matter~\cite{DarkMatter}, and so on. 
Models with extra dimensions should be regarded as effective theories 
with cut-off energy scales because they are nonrenormalizable and 
perturbative calculations will be invalid near those scales. 
Therefore it is important to estimate the cut-off scale of the model 
when we consider an extra-dimensional model. 
Tree-level unitarity provides a criterion for the perturbativity 
of a model at a given energy scale. 

The tree-level unitarity is usually discussed by evaluating 
scattering amplitudes of the longitudinally polarized 
weak bosons~$W_L^\pm$ and $Z_L$ at tree-level 
because they provide severer unitarity bound 
than other scattering processes. 
In SM, the Higgs boson plays an important role 
for the recovery of the unitarity. 
If it is sufficiently heavy and decoupled, the scattering amplitudes 
grow as $E^2$, where $E$ is the scattering energy, 
and exceed the unitarity bound at $\cO(1~\mbox{TeV})$. 
This means that perturbative calculations are no longer reliable 
above the scale. 
In the five-dimensional (5D) Higgsless models~\cite{Higgsless}, 
the tree-level unitarity is recovered by the Kaluza-Klein (KK) excitation 
modes of the gauge bosons instead of the Higgs boson in SM, 
and the unitarity violation delays up to $\cO(10~\mbox{TeV})$ 
when the compactification scale is assumed to be around 1~TeV. 

The situation is more complicated 
in the gauge-Higgs unification models~\cite{GHU1}-\cite{GHU5} 
because they have the Higgs mode as well as the KK gauge bosons, 
both of which participate in the unitarization of the theory. 
The gauge-Higgs unification is an attractive scenario as a solution to 
the gauge hierarchy problem. 
Higher dimensional gauge symmetry protects the electroweak scale 
against quantum corrections. 
The Higgs boson whose vacuum expectation value (VEV) breaks 
the electroweak gauge symmetry is identified with one of extra-dimensional 
components of the higher dimensional gauge fields, which we refer to as 
the gauge-scalars in this paper. 
The electroweak symmetry breaking is characterized 
by the Wilson line phase~$\thH$ along the extra dimension, 
which is gauge invariant. 
In these models, coupling constants and 
the KK mass scale~$\mKK$ depend on $\thH$ 
when we fix the W boson mass~$m_W$, and thus the scattering amplitudes 
for the weak bosons have nontrivial $\thH$-dependence. 
In particular in the models on the warped 
spacetime~\cite{warpGHU}-\cite{GHU:HS2}, 
the $WWH$ and $ZZH$ couplings ($H$ stands for the Higgs mode) deviate 
from the SM values and vanish at some specific values of $\thH$, 
such as $\pi$ or $\pi/2$, depending on the models~\cite{GHU:HS1,GHU:HS2}. 
For such values of $\thH$, the Higgs mode cannot participate in 
the unitarization of the weak boson scattering, and the amplitudes 
grow until the KK gauge bosons start to propagate and 
unitarize the scattering processes. 
Therefore it is important to understand the $\thH$-dependence of 
the scattering amplitudes for the weak bosons in order to estimate 
the unitarity violation scale~$\Lmd_{\rm uni}$. 
This issue is discussed in Ref.~\cite{FPR} and some qualitative behaviors 
of the amplitudes are clarified. 

In our previous work~\cite{HSY}, we investigated it 
more quantitatively by numerical calculations of 
the scattering amplitude for the process:~$W_L^++W_L^-\to Z_L+Z_L$ 
in the 5D $SU(3)$ gauge-Higgs unification model 
both in the flat and warped spacetimes. 
We found that the amplitude is enhanced for $\thH=\cO(1)$ in the warped case, 
which implies that the tree-level unitarity will be violated 
at a lower scale than that in the flat case. 
Although this result is expected to be common to 
the gauge-Higgs unification models, a specific value of $\Lmd_{\rm uni}$ 
depends on the model. 
It is well-known that the $SU(3)$ model is not realistic 
because it gives a wrong value of the Weinberg angle~$\tht_W$, \ie, 
$\sin^2\tht_W=3/4$. 
It is most interesting and useful to estimate $\Lmd_{\rm uni}$ 
in a realistic model, such as the 5D $SO(5)\times\ubl$ model, 
which was first proposed in Ref.~\cite{SO5model}. 

In this paper, we consider scattering of $W_L^+$ and $W_L^-$,  
investigate the $\thH$-dependence of the amplitudes, 
and numerically estimate $\Lmd_{\rm uni}$ 
in the 5D $SO(5)\times\ubl$ model. 
Note that $\thH$ and the Higgs mass $m_H$ are dynamically determined 
by quantum effect once the whole field content of the model is given. 
In the following discussion, however, we do not specify the fermion sector 
and treat $\thH$ and $m_H$ just as free parameters 
because we are interested in the tree-level amplitudes. 
These parameters parameterize the radiatively induced effective potential 
in a model-independent way. 
We take into account not only the elastic scattering but also 
possible inelastic scattering to obtain a proper unitarity bound.\footnote{
We do not consider inelastic scattering to fermions in the final states 
since we do not specify the fermion sector. 
Thus the bound~$\Lmd_{\rm uni}$ estimated here is a conservative one.  
} 
For the tree-level S-wave amplitude for the elastic scattering of the W bosons, 
there is an infrared divergence originating 
a singularity at forward scattering. 
We show an appropriate treatment to regularize this divergence 
by taking into account the instability of the W bosons in the final state. 


The paper is organized as follows. 
In Sec.~\ref{model}, we briefly review the $SO(5)\times\ubl$ gauge-Higgs 
unification model and provide necessary ingredients to calculate 
the scattering amplitudes for the weak bosons, 
which are extended versions 
of those used in Ref.~\cite{HSY} for the $SU(3)$ model. 
In Sec.~\ref{Wscattering}, we provide explicit expressions 
of the scattering amplitudes 
and show their behaviors as functions of $E$ and $\thH$ 
in the flat and warped spacetimes. 
In Sec.~\ref{unit_bound}, we estimate $\Lmd_{\rm uni}$ from 
the unitarity condition by using the amplitudes calculated 
in Sec.~\ref{Wscattering}. 
Sec.~\ref{summary} is devoted to the summary. 
In Appendix~\ref{basis_fcn}, we give definitions and explicit forms of 
the basis functions used in the text. 
In Appendix~\ref{5Dpropagator}, we derive the 5D propagators of 
the gauge fields. 
In Appendix~\ref{fwd_singularity}, we show a treatment of the singularity 
of the elastic scattering amplitude at forward scattering.

\section{{\boldmath $SO(5)\times\ubl$} model} \label{model}
In this section, we review the $SO(5)\times\ubl$ gauge-Higgs unification 
model~\cite{SO5model}. 
Most results in this section have been already obtained in the literature 
(see Ref.~\cite{GHU:HS2}, for example), 
but we repeat the discussion 
to explain our notation and for later convenience. 
 
\subsection{Set-up}
We consider an $SO(5)\times U(1)_X$ gauge theory compactified on $S^1/Z_2$. 
Arbitrary background metric with four-dimensional (4D) Poincar\'{e} symmetry 
can be written as 
\be
 ds^2 = G_{MN}dx^Mdx^N 
 = e^{-2\sgm(y)}\eta_{\mu\nu}dx^\mu dx^\nu+dy^2, 
\ee
where $M,N=0,1,2,3,4$ are 5D indices and $\eta_{\mu\nu}=\diag(-1,1,1,1)$. 
The fundamental region of $S^1/Z_2$ is $0\leq y\leq L$.  
The function $e^{\sgm(y)}$ is a warp factor, which is normalized 
as $\sgm(0)=0$. 
For example, $\sgm(y)=0$ in the flat spacetime, and $\sgm(y)=ky$ 
($0\leq y\leq L$) in the Randall-Sundrum warped spacetime~\cite{RS}, 
where $k$ is the inverse AdS curvature radius. 

The model has an $SO(5)$ gauge field~$A_M$ 
and a $\ubl$ gauge field~$B_M$. 
The former are decomposed as 
\be
 A_M = \sum_{\alp=1}^{10} A_M^\alp T^\alp
 = \sum_{\aL=1}^3A_M^{\aL}T^{\aL}+\sum_{\aR=1}^3A_M^{\aR}T^{\aR}
 +\sum_{\ha=1}^4A_M^{\ha}T^{\ha}, 
\ee
where $T^{\aL,\aR}$ ($\aL,\aR=1,2,3$) and $T^{\hat{a}}$ 
($\hat{a}=1,2,3,4$) are the generators of $SO(4)\sim \suL\times\suR$ 
and $SO(5)/SO(4)$, respectively, and are normalized as 
\be
 \tr(T^\alp T^\bt) = \frac{1}{2}\dlt^{\alp\bt}. 
\ee 
The 5D Lagrangian is 
\bea
 \cL \eql \sqrt{-G}\left[-\tr\brc{
 \frac{1}{2}G^{ML}G^{NP}F_{MN}^{(A)}F_{LP}^{(A)}
 +\frac{1}{\xi}\brkt{f_{\rm gf}^{(A)}}^2} \right. \nonumber\\
 &&\hspace{10mm}\left.
 -\brc{\frac{1}{4}G^{ML}G^{NP}F_{MN}^{(B)}F_{LP}^{(B)}
 +\frac{1}{2\xi}\brkt{f_{\rm gf}^{(B)}}^2}+\cdots \right], 
\eea
where $\sqrt{-G}\equiv\sqrt{-\det(G_{MN})}=e^{-4\sgm}$, 
$F_{MN}^{(A)}\equiv\der_M A_N-\der_N A_M-ig_A[A_M,A_N]$ 
($g_A$ is the 5D gauge coupling constant for $SO(5)$), 
$F_{MN}^{(B)}\equiv\der_M B_N-\der_N B_M$, 
and $\xi$ is a dimensionless parameter. 
The ellipsis denotes the ghost and the matter sectors, 
which are irrelevant to the following discussion. 
The gauge-fixing function~$f_{\rm gf}^{(A,B)}$ are chosen as 
\bea
 f_{\rm gf}^{(A)} \eql e^{2\sgm}\brc{\eta^{\mu\nu}\der_\mu A_\nu
 +\xi\cD_y^{\rm c}(e^{-2\sgm}A_y)}, \nonumber\\
 \cD_y^{\rm c}A_M \defa \der_y A_M-ig_A\sbk{A_y^{\rm bg},A_M}, \nonumber\\
 f_{\rm gf}^{(B)} \eql e^{2\sgm}\brc{\eta^{\mu\nu}\der_\mu B_\nu
 +\xi\der_y(e^{-2\sgm}B_y)}, 
\eea
where $A_y^{\rm bg}(y)$ is the classical background of $A_y(x,y)$. 

The boundary conditions for the gauge fields are written as 
\bea
 \begin{pmatrix} A_\mu \\ A_y \end{pmatrix}(x,y_i-y) 
 \eql Q_i\begin{pmatrix} A_\mu \\ -A_y \end{pmatrix}(x,y_i+y) Q_i^{-1}, 
 \nonumber\\
 \begin{pmatrix} B_\mu \\ B_y \end{pmatrix}(x,y_i-y)
 \eql \begin{pmatrix} B_\mu \\ -B_y \end{pmatrix}(x,y_i+y), 
\eea
where $i=0,L$, $y_0=0$, $y_L=L$, and $Q_i\in SO(5)$ are constant matrices 
satisfying $Q_i^2=1$. 
In the present paper we take $Q_0=Q_L=\diag(1,1,-1,-1)$
in the spinorial representation, or equivalently 
$Q_0=Q_L=\diag(-1,-1,-1,-1,1)$ in the vectorial representation.  
Then the gauge symmetry is broken to $SO(4)\times\ubl$ at both boundaries. 

We assume that the residual $SO(4)\times\ubl\sim\suL\times\suR\times\ubl$ 
is spontaneously broken to $\suL\times U(1)_Y$ at $y=0$ 
by some dynamics on the boundary, which leads to 
the following boundary mass terms.  
\be
 \cL_{\rm bd} = 2\sqrt{-g}\sbk{-\frac{M_\pm}{2}
 g^{\mu\nu}\brkt{A^{1_{\rm R}}_\mu A^{1_{\rm R}}_\nu
 +A^{2_{\rm R}}_\mu A^{2_{\rm R}}_\nu}
 -\frac{M_0}{2}g^{\mu\nu}A^{3'_{\rm R}}_\mu A^{3'_{\rm R}}_\nu}\dlt(y)+\cdots, 
 \label{L_bd}
\ee
where $g_{\mu\nu}=e^{-2\sgm}\eta_{\mu\nu}$, 
$\sqrt{-g}\equiv\sqrt{-\det(g_{\mu\nu})}=e^{-4\sgm}$, 
$M_{\pm}$ and $M_0$ are boundary mass parameters, and 
\be
 \begin{pmatrix} A_M^{3'_{\rm R}} \\ A_M^Y \end{pmatrix}
 \equiv \begin{pmatrix} c_\phi & -s_\phi \\ s_\phi & c_\phi \end{pmatrix}
 \begin{pmatrix} A_M^{3_{\rm R}} \\ B_M \end{pmatrix}, 
\ee
with 
\be
 c_\phi \equiv \frac{g_A}{\sqrt{g_A^2+g_B^2}}, \;\;\;\;\;
 s_\phi \equiv \frac{g_B}{\sqrt{g_A^2+g_B^2}}. 
\ee
Here $g_B$ is the 5D gauge coupling constant for $\ubl$. 
The gauge symmetry broken by these boundary mass terms can be recovered 
nonlinearly by introducing the Nambu-Goldstone (NG) modes 
localized at $y=0$. 

We do not specify the origin of the mass terms~(\ref{L_bd}) 
because it is irrelevant to the low-energy physics. 
We just assume that these masses are sufficiently heavier 
than the compactification scale. 
Then the boundary conditions for $A_\mu^{1_{\rm R}}$, 
$A_\mu^{2_{\rm R}}$ and $A_\mu^{3'_{\rm R}}$ at $y=0$ are effectively changed 
from the Neumann-type to the Dirichlet-type. 
In such a case, those for 
the gauge-scalars~$A_y^{1_{\rm R}}$, $A_y^{2_{\rm R}}$ and $A_y^{3'_{\rm R}}$ 
correspondingly change from Dirichlet to Neumann. 
The boundary degrees of freedom for the gauge-scalars 
at $y=0$ are provided by 
the boundary NG modes.\footnote{
The equations of motion for the boundary NG modes relates them to 
the boundary values of the gauge-scalars. }
As a result, the effective boundary conditions for the gauge fields are 
tabulated in Table~\ref{bdcd_gauge_fields}. 
\begin{table}[t]
\begin{center}
\begin{tabular}{|c|c|c|c|c|} \hline
\rule[-2mm]{0mm}{7mm}$A_\mu^{\aL}$ & $A_\mu^{1,2_{\rm R}}$ 
& $A_\mu^{3'_{\rm R}}$ & $A_\mu^Y$ & $A_\mu^{\ha}$ \\ \hline
(N,N) & (D,N) & (D,N) & (N,N) & (D,D) \\ \hline\hline
\rule[-2mm]{0mm}{7mm}$A_y^{\aL}$ & $A_y^{1,2_{\rm R}}$ & $A_y^{3'_{\rm R}}$ & 
$A_y^Y$ & $A_y^{\ha}$ \\ \hline
(D,D) & (N,D) & (N,D) & (D,D) & (N,N) \\ \hline
\end{tabular}
\end{center}
\caption{Boundary conditions for the gauge fields. 
The notation (D,N), for example, denotes the Dirichlet boundary condition 
at $y=0$ and the Neumann boundary condition at $y=L$. }
\label{bdcd_gauge_fields}
\end{table}

Note that only $(N,N)$ fields can have massless modes when perturbation theory 
is developed around the trivial configuration $A_M=B_M=0$. 
Thus the gauge symmetry is broken to $\suL\times U(1)_Y$ at tree-level. 
The zero-modes of the gauge-scalars form an $SU(2)$-doublet 
4D scalar~$(A_y^{\hat{1}}+iA_y^{\hat{2}},A_y^{\hat{4}}-iA_y^{\hat{3}})$, 
which plays a role of the Higgs doublet in SM 
whose VEV breaks $\suL\times U(1)_Y$ to the electromagnetic symmetry~$\uem$. 
They yield non-Abelian Aharonov-Bohm phases (Wilson line phases) 
when integrated along the fifth dimension. 
By using the residual $\suL\times\uy$ symmetry, we can always push 
the nonvanishing VEV into one component, say, $A_y^{\hat{4}}$. 
Then the Wilson line phase~$\thH$ is given by 
\be
 \thH =\frac{g_A}{\sqrt{2}}\int_0^L\dr y\;A_y^{\rm bg\,\hat{4}}(y). 
\ee

According to the transformation properties under the unbroken $\uem$ and 
the rotation by a constant matrix~$\Omg(L)$, the gauge fields are classified 
into the charged sector~$(A_M^{\pm_{\rm L}},A_M^{\pm_{\rm R}},A_M^{\hat{\pm}})
\equiv (A_M^{1_{\rm L}}\pm iA_M^{2_{\rm L}},A_M^{1_{\rm R}}\pm iA_M^{2_{\rm R}},
A_M^{\hat{1}}\pm iA_M^{\hat{2}})/\sqrt{2}$, the neutral sector 
$(A_M^{3_{\rm L}},A_M^{3_{\rm R}},B_M,A_M^{\hat{3}})$, 
and the ``Higgs'' sector~$A_M^{\hat{4}}$. 
Thus, in the following, we will use the index~$I$ 
which run over both the $SO(5)$-part~$\alp=(\aL,\aR,\ha)$ 
and the $U(1)$-part as 
\be
 I = I_+,I_-,I_0,\hat{4},  \label{index_I}
\ee
where $I_\pm = \pm_{\rm L},\pm_{\rm R},\hat{\pm}$ and 
$I_0 = 3_{\rm L},3_{\rm R},B,\hat{3}$.  
Then all the gauge fields are expressed in a matrix notation as 
\be
 \bdm{A}_M \equiv \sum_I A_M^IT^I, 
\ee
where $A_M^B\equiv B_M$. 
The generators are defined as 
\bea
 T^{\pm_{\rm L}} \defa \frac{1}{\sqrt{2}}\brkt{T^{1_{\rm L}}\mp iT^{2_{\rm L}}}, 
 \;\;\;\;\; 
 T^{\pm_{\rm R}} \equiv \frac{1}{\sqrt{2}}\brkt{T^{1_{\rm R}}\mp iT^{2_{\rm R}}},
 \nonumber\\
 T^{\hat{\pm}} \defa \frac{1}{\sqrt{2}}\brkt{T^{\hat{1}}\mp iT^{\hat{2}}}, 
 \;\;\;\;\;
 T^B \equiv \frac{1}{\sqrt{2d}}\bdm{1_d},  
\eea
where $d$ is a dimension of the representation. 
The structure constants in this basis are listed in Table~\ref{str_cst}. 
\begin{table}[t]
\begin{center}
\begin{tabular}{ccc|c||ccc|c||ccc|c}
 \hline \rule[-2mm]{0mm}{7mm} $I$ & $J$ & $K$ & $C^{IJK}$ & 
 $I$ & $J$ & $K$ & $C^{IJK}$ &
 $I$ & $J$ & $K$ & $C^{IJK}$ \\ \hline
 \rule[-2mm]{0mm}{7mm} $+_{\rm L}$ & $-_{\rm L}$ & $3_{\rm L}$ & $i$ 
 & $3_{\rm L}$ & $\hat{+}$ & $\hat{-}$ & $i/2$ 
 & $-_{\rm R}$ & $\hat{+}$ & $\hat{3}$ & $-i/2$ \\ 
 \rule[-2mm]{0mm}{7mm} $+_{\rm L}$ & $\hat{-}$ & $\hat{3}$ & $i/2$ 
 & $3_{\rm L}$ & $\hat{3}$ & $\hat{4}$ & $1/2$ 
 & $-_{\rm R}$ & $\hat{+}$ & $\hat{4}$ & $-1/2$ \\ 
 \rule[-2mm]{0mm}{7mm} $+_{\rm L}$ & $\hat{-}$ & $\hat{4}$ & $1/2$ 
 & $+_{\rm R}$ & $-_{\rm R}$ & $3_{\rm R}$ & $i$ 
 & $3_{\rm R}$ & $\hat{+}$ & $\hat{-}$ & $i/2$ \\ 
 \rule[-2mm]{0mm}{7mm} $-_{\rm L}$ & $\hat{+}$ & $\hat{3}$ & $-i/2$ 
 & $+_{\rm R}$ & $\hat{-}$ & $\hat{3}$ & $i/2$ 
 & $3_{\rm R}$ & $\hat{3}$ & $\hat{4}$ & $-1/2$ \\ 
 \rule[-2mm]{0mm}{7mm} $-_{\rm L}$ & $\hat{+}$ & $\hat{4}$ & $1/2$ 
 & $+_{\rm R}$ & $\hat{-}$ & $\hat{4}$ & $-1/2$ & & & \\
 \hline
\end{tabular}
\end{center}
\caption{The structure constants for the generators~$T^I$. 
For the other combinations of indices, $C^{IJK}=0$. }  
\label{str_cst}
\end{table}
The orthonormal conditions for the generators are written as 
\be
 \tr\brkt{T^IT^{\bar{J}}} = \frac{1}{2}\dlt^{I\bar{J}}, 
\ee
where the index~$\bar{J}$ runs as 
\be
 \bar{J} = J_-,J_+,J_0,\hat{4}. 
 \label{index_barJ}
\ee

\subsection{Mode expansion}
The expansion of the 5D gauge fields into 4D KK modes is performed 
in a conventional way (see Ref.~\cite{GHU:HS2}, for example). 
We move to the Scherk-Schwarz basis, in which $\tl{A}_y^{\rm bg}=0$. 
It is related to the original basis by the gauge transformation, 
\bea
 \bdm{\tl{A}}_M \eql \Omg \bdm{A}_M \Omg^{-1}
 -\frac{i}{g_A}(\der_M\Omg)\Omg^{-1},  \label{gauge_trf}
\eea
with
\be
 \Omg(y) \equiv \cP\exp\brc{-ig_A\int_0^y\dr y'\;
 A_y^{\rm bg\, \hat{4}}(y')\, T^{\hat{4}}}. \label{def:Omg}
\ee
The symbol~$\cP$ stands for the path-ordered operator from left to right. 

For the following discussion, it is convenient to move to the momentum 
representation for the 4D part while remain the coordinate representation 
for the fifth dimension~\cite{GP}. 
Then the 5D gauge fields are expanded into the KK modes as 
\bea
 \tl{A}_\mu^I(p,y) \eql \sum_n u_n^I(y)A_\mu^{(n)}(p)
 +\sum_n w_n^I(y)p_\mu A_{\rm S}^{(n)}(p), \nonumber\\
 \tl{A}_y^I(p,y) \eql \sum_n v_n^I(y)\vph^{(n)}(p). 
\eea
Notice that $\tl{A}_\mu^I(p,y)$ are decomposed into two parts, 
according to their polarization. 
In the above expression, $A_\mu^{(n)}(p)$ are polarized as 
$p^\mu A_\mu^{(n)}(p)=0$ and include the transverse and the longitudinal 
modes, which are physical for the massive modes. 
On the other hand, $A_{\rm S}^{(n)}(p)$ are unphysical scalar modes. 
The gauge-scalar modes~$\vph^{(n)}(p)$ are also unphysical 
besides the zero-mode. 

By solving the mode equations 
with the boundary conditions shown in Table~\ref{bdcd_gauge_fields}, 
the mode functions are expressed by the basis functions~$C_0(y,m)$ and 
$S_0(y,m)$ defined in Appendix~\ref{basis_fcn} in the vector notation for 
the index~$I$ as  
\bea
 \vec{u}_n(y) \eql \cM_0(y,m_n)\vec{N}_n, \nonumber\\
 \vec{w}_n(y) \eql \cM_0(y,\tl{m}_n/\sqrt{\xi})\vec{\tl{N}}_n, \nonumber\\
 \vec{v}_n(y) \eql \frac{d}{dy}\vec{w}_n(y), 
\eea
where the matrix~$\cM_0(y,m)$ is a function defined 
by Eqs.(\ref{def:cM}) and (\ref{def:cMs}), 
$m_n$ is a mass eigenvalue for $A_\mu^{(n)}$, and $\tl{m}_n$ is 
a common mass eigenvalue for $A_{\rm S}^{(n)}$ and $\vph^{(n)}$. 
The constant vectors~$\vec{N}_n$, $\vec{\tl{N}}_n$ are determined by 
\be
 \cW(m_n)\vec{N}_n = 0, \;\;\;\;\;
 \cW(\tl{m}_n)\vec{\tl{N}}_n = 0,  \label{cW_N}
\ee
where the matrix~$\cW(m)$ is defined by Eq.(\ref{def:cW}), 
and by the orthonormal conditions
\bea
 \int_0^L\dr y\;\vec{u}_m(y)\cdot\vec{u}_n(y) \eql \dlt_{mn}, \nonumber\\
 \frac{\tl{m}_n^2}{\xi}\int_0^L\dr y\;\vec{w}_m(y)\cdot\vec{w}_n(y) 
 \eql \int_0^L\dr y\;e^{-2\sgm(y)}\vec{v}_m(y)\cdot\vec{v}_n(y) 
 = \dlt_{mn}. \label{orthonormal}
\eea
The conditions that Eq.(\ref{cW_N}) has nontrivial solutions are 
\be
 \det\cW(m_n) = 0, \;\;\;\;\;
 \det\cW(\tl{m}_n) = 0, 
\ee
which determine the mass eigenvalues~$m_n$ and $\tl{m}_n$. 

Here we give explicit expressions of light modes. 
The $W$ boson is identified with the lightest mode 
in the charged sector. 
Its mass~$m_W$ is determined as the lowest solution to 
\be
 C'_0(L,m_W)S_0(L,m_W)+\frac{m_We^{\sgm(L)}}{2}\sin^2\thH = 0, 
\ee
and the corresponding mode function is calculated as 
\be
 u_W^{I_+}(y) 
 = \sum_{J_+}\cM_0^{{\rm ch}\,I_+J_+}(y,m_W)N_W^{J_+}, \;\;\;\;\;
 u_W^{I_-}(y) = u_W^{I_0}(y) = u_W^{\hat{4}}(y) = 0, 
 \label{def:uW}
\ee
for the $W^+$ boson, and 
\be
 u_W^{I_-}(y) 
 = \sum_{J_-}\cM_0^{{\rm ch}\,I_-J_-}(y,m_W)N_W^{J_-}, \;\;\;\;\;
 u_W^{I_+}(y) = u_W^{I_0}(y) = u_W^{\hat{4}}(y) = 0, 
 \label{def:uW2}
\ee
for the $W^-$ boson. 
Here $\cM_0^{\rm ch}(y,m)$ is defined in Eq.(\ref{def:cMs}) and 
\be
 N_W = \alp_W\brkt{-S'_0(L,m_W),C'_0(L,m_W),\sqrt{2}C'_0(L,m_W)\cot\thH}^t. 
\ee
The constant $\alp_W$ is determined 
by the normalization condition~(\ref{orthonormal}). 

The neutral sector has a zero-mode, which corresponds to the photon. 
Its mode function is a constant vector, 
\be
 u_\gm^{I_0}(y) = \sqrt{\frac{1}{(1+s_\phi^2)L}}
 (s_\phi,s_\phi,c_\phi,0), \;\;\;\;\;
 u_\gm^{I_\pm}(y) = u_\gm^{\hat{4}}(y) = 0. 
\ee
The $Z$ boson is identified with the second lightest mode 
in the neutral sector. 
Its mass~$m_Z$ is determined as the lowest solution to 
\be
 C'_0(L,m_Z)S_0(L,m_Z)+\frac{m_Ze^{\sgm(L)}(1+s_\phi^2)}{2}\sin^2\thH = 0, 
\ee
and the corresponding mode function is 
\be
 u_Z^{I_0}(y) = \sum_{J_0}\cM_0^{{\rm nt}\,I_0J_0}(y,m_Z)N_Z^{J_0}, \;\;\;\;\;
 u_Z^{I_\pm}(y) = u_Z^{\hat{4}}(y) = 0, 
 \label{def:uZ}
\ee
where $\cM_0^{\rm nt}(y,m)$ is defined in Eq.(\ref{def:cMs}) and 
\be
 N_Z = \alp_Z\brkt{-S'_0,c_\phi^2C'_0+s_\phi^2S'_0,
 s_\phi c_\phi(S'_0-C'_0),\sqrt{2}C'_0\cot\thH}^t. 
\ee
The arguments in the right-hand side are $(L,m_Z)$, and 
the normalization constant $\alp_Z$ is determined by Eq.(\ref{orthonormal}). 

In the Randall-Sundrum spacetime ($\sgm(y)=ky$), the basis functions are 
expressed by the Bessel functions as shown in Appendix~\ref{basis_fcn}. 
When the warp factor~$e^{\sgm(L)}=e^{kL}$ is large enough, 
the masses of the $W$ and $Z$ bosons are approximated as 
\be
 m_W\simeq \frac{\mKK}{\pi}\sqrt{\frac{1}{kL}}\abs{\sin\thH}, \;\;\;\;\;
 m_Z\simeq \frac{\mKK}{\pi}\sqrt{\frac{1+s_\phi^2}{kL}}\abs{\sin\thH}, 
 \label{mWZ:warp}
\ee
where 
\be
 \mKK\equiv\frac{k\pi}{e^{kL}-1}
\ee
is the KK mass scale. 
Thus the Weinberg angle~$\tht_W$ is expressed in terms of $s_\phi$ as 
\be
 \tan\tht_W \simeq s_\phi. 
\ee

In the flat spacetime ($\sgm(y)=0$), the $W$ and $Z$ boson masses are 
expressed as 
\be
 m_W = \frac{1}{L}\sin^{-1}\brkt{\frac{1}{\sqrt{2}}\sin\thH}, \;\;\;\;\;
 m_Z = \frac{1}{L}\sin^{-1}\brkt{\sqrt{\frac{1+s_\phi^2}{2}}\sin\thH}. 
 \label{mWZ:flat}
\ee
In contrast to the $SU(3)$ model, the spectrum is not linear 
for the ``Higgs VEV''~$\thH$ 
even in the flat spacetime~\cite{GHU:HS2}. 
This stems from the fact that the mechanism of mass generation 
for the 4D gauge bosons involves not only 4D gauge fields in each KK level, 
but also fields in other KK levels. 
In the original basis, the $W$ boson mass term comes from 
\bea
 \cL \eql g_A^2e^{-2\sgm}\eta^{\mu\nu}\tr\brc{\sbk{A_\mu,A_y}\sbk{A_\nu,A_y}}
 +\cdots \nonumber\\
 \eql -\frac{g_A^2e^{-2\sgm}}{8}\brkt{A_y^{\hat{4}}}^2
 \eta^{\mu\nu}\brkt{A^{+_{\rm L}}_\mu A^{-_{\rm L}}_\nu
 -A^{+_{\rm L}}_\mu A^{-_{\rm R}}_\nu-A^{+_{\rm R}}_\mu A^{-_{\rm L}}_\nu
 +A^{+_{\rm R}}_\mu A^{-_{\rm R}}_\nu}+\cdots. 
 \label{L_mass}
\eea
In the flat spacetime, the profile of $A_y^{{\rm bg}\,\hat{4}}$ is flat. 
Thus there would be no mixing among different KK levels due to 
the orthogonality of the mode functions if the mixing terms between 
the $\suL$ and $\suR$ gauge fields were absent in Eq.(\ref{L_mass}), 
just like the case of the $SU(3)$ model. 
However, the KK level mixing actually occurs 
due to the presence of the mixing terms between the $\suL$ and $\suR$ 
gauge fields whose boundary conditions are different 
(see Table~\ref{bdcd_gauge_fields}). 
Then the lowest mode in each KK tower necessarily mixes with heavy KK modes 
when $A_y^{\hat{4}}$, or $\thH$, acquires a nonzero value. 
This mixing makes the $\thH$-dependence of the spectrum nonlinear.

\subsection{5D propagators} \label{5D_propagators}
For the purpose of calculating the scattering amplitude, it is convenient 
to use the 5D propagators~$G_{\rm T}(y,y',\sqrt{-p^2})$ 
defined in a mixed momentum/position representation~\cite{GP}. 
It describes the propagation of the entire KK towers of excitations 
carrying the 4D momentum~$p$ between two points $y$ and $y'$ 
in the extra dimension. 
This approach has an advantage that we need not explicitly calculate 
mass eigenvalues and mode functions 
for modes propagating in the internal lines of the Feynmann diagrams, 
nor sum over contributions from 
infinite (or large) number of KK modes.\footnote{
This approach is also useful for models with continuum spectra~\cite{Unhiggs}.
}
The definition and the derivation of the 5D propagator are given 
in Appendix~\ref{5Dpropagator}. 
It is expressed from Eq.(\ref{expr:G_T}) in the following block-diagonal form. 
\be
 G_{\rm T} = \begin{pmatrix} G_{\rm T}^{\rm ch} & & & \\
 & G_{\rm T}^{\rm ch} & & \\ & & G_{\rm T}^{\rm nt} & \\
 & & & G_{\rm T}^{\hat{4}\hat{4}} \end{pmatrix}, 
\ee
where
\bea
 G_{\rm T<}^{\rm ch}(y,y',\abs{p}) \eql 
 e^{2\sgm(L)}\cM_0^{\rm ch}(y,\abs{p})
 \cW_{\rm ch}^{-1}(\abs{p})\cM_L^{\rm ch}(y',\abs{p})R_\tht^{\rm ch}, 
 \nonumber\\
 G_{\rm T<}^{\rm nt}(y,y',\abs{p}) \eql 
 e^{2\sgm(L)}\cM_0^{\rm nt}(y,\abs{p})
 \cW_{\rm nt}^{-1}(\abs{p})\cM_L^{\rm nt}(y',\abs{p})R_\tht^{\rm nt}, 
 \nonumber\\
 G_{\rm T<}^{\hat{4}\hat{4}}(y,y',\abs{p}) \eql 
 \frac{e^{2\sgm(L)}S_0(y,\abs{p})S_L(y',\abs{p})}{\abs{p}S_0(L,\abs{p})},  
\eea
and $\abs{p}\equiv\sqrt{-p^2}$. 
The explicit forms of the matrices in the right-hand sides are given 
in Appendix~\ref{5Dpropagator}. 

Using the mode equation and Eq.(\ref{eq_propagator}) 
with the boundary conditions, we can show the following relation. 
\be
 \vec{u}_n(y) = -(p^2+m_n^2)\int_0^L\dr y'\;G_{\rm T}(y,y',\abs{p})\vec{u}_n(y'). 
\ee
Thus the 5D propagator can also be expressed as 
\be
 G_{\rm T}(y,y',\abs{p}) = -\sum_n\frac{\vec{u}_n(y)\vec{u}_n^{\,t}(y')}
 {p^2+m_n^2}.  \label{KKexpand_propagator}
\ee

\section{Weak boson scattering} \label{Wscattering}
Now we consider the scattering of the weak bosons. 
The scattering amplitudes are functions of the total energy~$E$ 
and the scattering angle~$\chi$ in the center-of-mass frame. 
Let us consider the scattering process: 
$\ket{p_1,\vep_1,m}+\ket{p_2,\vep_2,n}\to\ket{p_3,\vep_3,l}+\ket{p_4,\vep_4,k}$, 
where $p_i$ and $\vep_i$ ($i=1,2,3,4$) denote the 4-momenta and 
the polarization vectors respectively, and $m,n,\cdots$ labels the particle 
species including the KK levels.

\subsection{Scattering amplitudes}
As mentioned in Sec.~\ref{5D_propagators}, the scattering amplitudes are 
easily calculated by utilizing the 5D propagators. 
The tree-level amplitude~$\cA$ for the vector boson scattering is expressed by 
\be
 \cA = \cA^{\rm C}+\cA^{\rm V}+\cA^{\rm S}, 
\ee
where $\cA^{\rm C}$, $\cA^{\rm V}$ and $\cA^{\rm S}$ are contributions 
from the contact interactions, exchange of the vector modes 
and that of the gauge-scalar modes, respectively, and are given by 
\bea
 \cA^{\rm C}_{mnlk} \eql -ig_A^2\int_0^L\dr y\;\sum_I\left[
 \brc{U_{ml}^I(y)U_{nk}^{\bar{I}}(y)+U_{mk}^I(y)U_{nl}^{\bar{I}}(y)}
 (\vep_1\cdot\vep_2)(\vep_3^*\cdot\vep_4^*) \right. \nonumber\\
 &&\hspace{30mm}
 +\brc{U_{mn}^I(y)U_{lk}^{\bar{I}}(y)+U_{mk}^I(y)U_{ln}^{\bar{I}}(y)}
 (\vep_1\cdot\vep_3^*)(\vep_2\cdot\vep_4^*) \nonumber\\
 &&\hspace{30mm}\left.
 +\brc{U_{mn}^I(y)U_{kl}^{\bar{I}}(y)+U_{ml}^I(y)U_{kn}^{\bar{I}}(y)}
 (\vep_1\cdot\vep_4^*)(\vep_2\cdot\vep_3^*)\right], \label{cA^C} \\
 \cA^{\rm V}_{mnlk} \eql -ig_A^2\sum_{I,J}\int_0^L\dr y\int_0^L\dr y'\;
 U_{mn}^{\bar{I}}(y)G_{\rm T}^{I\bar{J}}(y,y',\abs{p_{12}})
 U_{lk}^{J}(y')P_{1234} \nonumber\\
 &&+ig_A^2\sum_{I,J}\int_0^L\dr y\int_0^L\dr y'\;
 U_{ml}^{\bar{I}}(y)G_{\rm T}^{I\bar{J}}(y,y',\abs{p_{13}})
 U_{nk}^{J}(y')P_{1324} \nonumber\\
 &&+ig_A^2\sum_{I,J}\int_0^L\dr y\int_0^L\dr y'\;
 U_{mk}^{\bar{I}}(y)G_{\rm T}^{I\bar{J}}(y,y',\abs{p_{14}})
 U_{nl}^{J}(y')P_{1423}, \label{cA^V} \\
 \cA^{\rm S}_{mnlk} \eql ig_A^2\sum_I\int_0^L\dr y\;e^{2\sgm(y)}
 \left\{Y_{mn}^I(y)Y_{lk}^{\bar{I}}(y)
 \frac{(\vep_1\cdot\vep_2)(\vep_3^*\cdot\vep_4^*)}{p_{12}^2} \right. \nonumber\\
 &&\hspace{10mm}\left. 
 +Y_{ml}^I(y)Y_{nk}^{\bar{I}}(y)
 \frac{(\vep_1\cdot\vep_3^*)(\vep_2\cdot\vep_4^*)}{p_{13}^2}
 +Y_{mk}^I(y)Y_{nl}^{\bar{I}}(y)
 \frac{(\vep_1\cdot\vep_4^*)(\vep_2\cdot\vep_3^*)}{p_{14}^2}\right\}, 
 \label{cA^S}
\eea
where $p_{12}\equiv p_1+p_2$, $p_{13}\equiv p_1-p_3$, $p_{14}\equiv p_1-p_4$, 
\bea
 P_{1234} \defa \brc{2(p_1\cdot\vep_2)\vep_1-2(p_2\cdot\vep_1)\vep_2
 -(\vep_1\cdot\vep_2)(p_1-p_2)}^\mu
 \brkt{\eta_{\mu\nu}-\frac{p_{12\mu}p_{12\nu}}{p_{12}^2}} \nonumber\\
 &&\times\brc{2(p_3^*\cdot\vep_4^*)\vep_3^*-2(p_4\cdot\vep_3^*)\vep_4^*
 -(\vep_3^*\cdot\vep_4^*)(p_3-p_4)}^\nu,  \nonumber\\
 P_{1324} \defa \brc{2(p_1\cdot\vep_3^*)\vep_1+2(p_3\cdot\vep_1)\vep_3^*
 -(\vep_1\cdot\vep_3^*)(p_1+p_3)}^\mu
 \brkt{\eta_{\mu\nu}-\frac{p_{13\mu}p_{13\nu}}{p_{13}^2}} \nonumber\\
 &&\times\brc{2(p_2\cdot\vep_4^*)\vep_2+2(p_4\cdot\vep_2)\vep_4^*
 -(\vep_2\cdot\vep_4^*)(p_2+p_4)}^\nu, \nonumber\\
 P_{1423} \defa \brc{2(p_1\cdot\vep_4^*)\vep_1+2(p_4\cdot\vep_1)\vep_4^*
 -(\vep_1\cdot\vep_4^*)(p_1+p_4)}^\mu
 \brkt{\eta_{\mu\nu}-\frac{p_{14\mu}p_{14\nu}}{p_{14}^2}} \nonumber\\
 &&\times\brc{2(p_2\cdot\vep_3^*)\vep_2+2(p_3\cdot\vep_2)\vep_3^*
 -(\vep_2\cdot\vep_3^*)(p_2+p_3)}^\nu, 
\eea
and the functions in the integrands are defined as 
\bea
 U_{mn}^I(y) \defa C^{IJK}u_m^J(y)u_n^K(y), \nonumber\\
 Y_{mn}^I(y) \defa e^{-2\sgm(y)}C^{IJK}
 \brc{\brkt{u_m^J}'(y)u_n^K(y)-u_m^J(y)\brkt{u_n^K}'(y)}.  
\eea
Here we have used the relation~$p_i\cdot\vep_i(p_i)=0$ ($i=1,2,3,4$). 
The prime denotes derivative with respect to $y$. 

The first, second and third lines in Eq.(\ref{cA^V}) correspond to 
the $s$-, $t$- and $u$-channel diagrams 
exchanging the 4D vector modes, respectively. 
The above expression of the amplitude is a result of a cancellation 
between the gauge-dependent part~$G_{\rm S}(y,y',\abs{p})$ 
in the propagator of the vector modes and 
the gauge-scalar propagator~$G_{yy}(y,y',\abs{p})$. 
This cancellation occurs due to the relation (\ref{rel:Gyy-Gs}) 
and makes the resultant amplitude gauge-independent. 
The contribution~$\cA^{\rm S}$ is a remnant of the cancellation. 

The gauge invariance of the theory ensures the equivalence theorem~\cite{ET}, 
which states that the scattering of the longitudinally polarized vector bosons 
is equivalent to that of the (would-be) NG bosons eaten by the gauge bosons. 
In 5D models, the gauge-scalar modes~$\vph^{(n)}$ coming from $A_y$ play 
the role of the NG bosons in the equivalence 
theorem~\cite{Higgsless,5D_ET}. 
Namely, the following relation holds for the longitudinal vector 
modes~$A_L^{(n)}$. 
\be
 T(A_L^{(n_1)},\cdots,A_L^{(n_l)};\Phi) 
 = C_l T(i\vph^{(n_1)},\cdots,i\vph^{(n_l)};\Phi)+\cO\brkt{\frac{M^2}{E^2}}, 
 \label{Eq_Th}
\ee
where all external lines are directed inwards, 
$\Phi$ denotes any possible amputated external physical fields, 
such as the transverse gauge boson, and $M$ is the heaviest mass 
among the external lines.  
A constant~$C_l$ is gauge-dependent, but $C_l=1$ at tree-level.\footnote{
We can also take a gauge where $C_l=1$ at all orders of the perturbative 
expansion~\cite{ET_loop}. 
}
The correction term is $\cO(M^2/E^2)$ 
because of the 5D gauge invariance (see Ref.~\cite{tanabashi}, for example). 
Eq.(\ref{Eq_Th}) is useful to discuss the high-energy behavior of 
the scattering amplitude~$\cA$ because the corresponding NG boson amplitude 
does not have $\cO(E^4)$ contributions,\footnote{
For the non-forward (non-backward) scattering, $\cO(E^2)$ contributions 
are also absent. } 
which makes it easier to numerically calculate the amplitude 
thanks to the absence of cancellations between large numbers. 

The scattering amplitude for the corresponding NG bosons comes only from 
diagrams exchanging the vector modes. 
\bea
 \cB_{mnlk} \eql -ig_5^2\sum_{I,J}\int_0^L\dr y\int_0^L\dr y'\;
 V_{mn}^{\bar{I}}(y)(p_1-p_2)^\mu G_{\mu\nu}^{I\bar{J}}(p_{12},y,y')
 (p_3-p_4)^\nu V_{lk}^{J}(y')  \nonumber\\
 &&+ig_5^2\sum_{I,J}\int_0^L\dr y\int_0^L\dr y'\;
 V_{ml}^{\bar{I}}(y)(p_1+p_3)^\mu G_{\mu\nu}^{I\bar{J}}(p_{13},y,y')
 (p_2+p_4)^\nu V_{nk}^{J}(y')  \nonumber\\
 &&+ig_5^2\sum_{I,J}\int_0^L\dr y\int_0^L\dr y'\;
 V_{mk}^{\bar{I}}(y)(p_1+p_4)^\mu G_{\mu\nu}^{I\bar{J}}(p_{14},y,y')
 (p_2+p_3)^\nu V_{nl}^{J}(y'), 
 \label{cB}
\eea
where
\bea
 V_{mn}^I(y) \defa e^{-2\sgm(y)}C^{IJK}v_m^J(y)v_n^K(y). 
\eea

\subsection{Various behaviors of the amplitudes}
Here we show various behaviors of the scattering amplitudes given 
in the previous subsection. 
For numerical calculation, we consider the flat ($\sgm(y)=0$) 
and the Randall-Sundrum ($\sgm(y)=ky$) spacetimes, 
and choose the gauge parameter as $\xi=1$, 
the 4D weak gauge coupling~$g\equiv g_A/\sqrt{L}$ 
as $g^2=4\pi\alp_{\rm EM}/\sin^2\tht_W=0.4$. 
We take the $W$ boson mass~$m_W$ as an input parameter. 
Then the size of the extra dimension~$L$ becomes $\thH$-dependent 
after fixing $m_W$. 
(See Eqs.(\ref{mWZ:warp}) and (\ref{mWZ:flat}).)
The KK mass scale~$m_{\rm KK}=\pi k/(e^{kL}-1)$ also depends on 
$\thH$ for a given value of the warp factor~$e^{kL}$. 
Thus the amplitudes are functions of the center-of-mass energy~$E$, 
the Wilson line phase~$\thH$ and the warp factor~$e^{kL}$.\footnote{
The Wilson line phase~$\thH$ is dynamically determined at quantum level 
if we fix the whole matter content of the model.} 
The physical amplitude~$\cA$ is of course gauge-independent, 
and the $\xi$-dependence of the gauge-scalar scattering amplitude~$\cB$ is small 
in high-energy region as can be seen from Eq.(\ref{Eq_Th}). 

Here let us comment on another advantage of using the 5D propagators. 
By using the relation~(\ref{KKexpand_propagator}), 
the scattering amplitudes given in the previous subsection are rewritten as 
more conventional forms in the KK analysis. 
For example, Eq.(\ref{cA^V}) is rewritten as 
\bea
 \cA^{\rm V}_{mnlk} \eql i\sum_r\brc{
 \frac{\lmd_{mnr}\lmd_{lkr}}{p_{12}^2+m_r^2}P_{1234}
 -\frac{\lmd_{mlr}\lmd_{nkr}}{p_{13}^2+m_r^2}P_{1324} 
 -\frac{\lmd_{mkr}\lmd_{nlr}}{p_{14}^2+m_r^2}P_{1423}}, 
 \label{cA^V2}
\eea
where 
\be
 \lmd_{mnr} \equiv g_A\int_0^L\dr y\;C^{IJK}u_r^I(y)u_m^J(y)u_n^K(y)
\ee
is a 4D effective coupling constant among the KK modes. 
Below $\mKK$, contributions of the heavy KK modes 
in the infinite sum are negligible because of suppression by 
large KK masses in the 4D propagators. 
Thus we can approximate $\cA^{\rm V}_{mnlk}$ in a good accuracy
by picking up only finite number of light modes in Eq.(\ref{cA^V2}). 
However, above $\mKK$, such an approximation becomes worse 
and much larger number of KK modes are necessary for summation 
in order to keep the accuracy of the approximation. 
Therefore this approximation is not practical for our purpose 
since we would like to see the behaviors of the scattering amplitudes 
beyond the KK mass scale. 
The 5D propagators enable us to calculate the amplitudes 
in high-energy region with sufficient accuracy. 

The 4 momenta and the polarization vectors of the initial and final 
states are parameterized as in Table~\ref{Ep}. 
\begin{table}[t]
\begin{center}
\begin{tabular}{|l|l|} \hline
$p_1=(E_1,0,0,p_I)$ & $\vep_1(p_1)=(p_I,0,0,E_1)/m_m$ \\
$p_2=(E_2,0,0,-p_I)$ & $\vep_2(p_2)=(p_I,0,0,-E_2)/m_n$ \\
$p_3=(E_3,p_F\sin\chi,0,p_F\cos\chi)$ & 
$\vep_3(p_3)=(p_F,E_3\sin\chi,0,E_3\cos\chi)/m_l$ \\
$p_4=(E_4,-p_F\sin\chi,0,-p_F\cos\chi)$ & 
$\vep_4(p_4)=(p_F,-E_4\sin\chi,0,-E_4\cos\chi)/m_k$ \\ \hline
\end{tabular}
\end{center}
\caption{The 4 momenta and the polarization vectors of the initial 
and the final states. 
The definitions of $E_i$ ($i=1,2,3,4$), $p_I$ and $p_F$ are given 
in Eq.(\ref{def:E_is}), and $\chi$ is the scattering angle 
in the center-of-mass frame. 
}
\label{Ep}
\end{table}
There, $\chi$ is the scattering angle in the center-of-mass frame, 
and the energy and the momentum of each particle are expressed as 
\bea
 E_1 \eql \frac{E}{2}+\frac{m_m^2-m_n^2}{2E}, \;\;\;\;\;
 E_2 = \frac{E}{2}+\frac{m_n^2-m_m^2}{2E}, \nonumber\\
 p_I \eql \sqrt{E_1^2-m_m^2} = \sqrt{E_2^2-m_n^2}
 = \brc{\frac{E^2}{4}-\frac{m_m^2+m_n^2}{2}+\frac{(m_m^2-m_n^2)^2}{4E^2}}^{1/2}, 
 \nonumber\\
 E_3 \eql \frac{E}{2}+\frac{m_l^2-m_k^2}{2E}, \;\;\;\;\;
 E_4 = \frac{E}{2}+\frac{m_k^2-m_l^2}{2E}, \nonumber\\
 p_F \eql \sqrt{E_3^2-m_l^2} = \sqrt{E_4^2-m_k^2}
 = \brc{\frac{E^2}{4}-\frac{m_l^2+m_k^2}{2}+\frac{(m_l^2-m_k^2)^2}{4E^2}}^{1/2}, 
 \label{def:E_is}
\eea
where $E$ is the total energy in the center-of-mass frame.

\subsubsection{Non-forward scattering}
First we consider the non-forward (and non-backward) scattering. 
We choose the scattering angle as $\chi=\pi/3$ in the following. 
Let us consider the process:~$W_L^++W_L^-\to Z_L+Z_L$, as an example. 
In this case, the mode functions in Eqs.(\ref{cA^C})-(\ref{cA^S}) are 
taken as 
\bea
 u_m \eql (u_W^{+_{\rm L}},u_W^{+_{\rm R}},u_W^{\hat{+}},0,0,0,0,0,0,0,0), 
 \nonumber\\
 u_n \eql (0,0,0,u_W^{-_{\rm L}},u_W^{-_{\rm R}},u_W^{\hat{-}},0,0,0,0,0), 
 \nonumber\\
 u_l \eql u_k = (0,0,0,0,0,0,u_Z^{3_{\rm L}},u_Z^{3_{\rm R}},u_Z^B,
 u_Z^{\hat{3}},0), 
\eea
where $u_W^{I_\pm}(y)$ and $u_Z^{I_0}(y)$ are defined 
in Eqs.(\ref{def:uW}), (\ref{def:uW2}) and (\ref{def:uZ}), respectively. 
Then, Eqs.(\ref{cA^C})-(\ref{cA^S}) are reduced to 
\bea
 \cA^{\rm C}_{WWZZ} \eql -ig_A^2\int_0^L\dr y\;U_{WZ}^2(y)
 \brc{2(\vep_1\cdot\vep_2)(\vep_3^*\cdot\vep_4^*)
 -(\vep_1\cdot\vep_3^*)(\vep_2\cdot\vep_4^*)
 -(\vep_1\cdot\vep_4^*)(\vep_2\cdot\vep_3^*)}, \nonumber\\
 \cA^{\rm V}_{WWZZ} \eql ig_A^2\int_0^L\dr y\int_0^L\dr y'\;
 U_{WZ}(y)\cdot G_{\rm T}^{\rm ch}(y,y',\abs{p_{13}})\cdot U_{WZ}(y')
 P_{1324} \nonumber\\
 &&+ig_A^2\int_0^L\dr y\int_0^L\dr y'\;
 U_{WZ}(y)\cdot G_{\rm T}^{\rm ch}(y,y',\abs{p_{14}})\cdot U_{WZ}(y')
 P_{1423}, \nonumber\\
 \cA^{\rm S}_{WWZZ} \eql ig_A^2\int_0^L\dr y\;e^{2\sgm(y)}\left[
 Y_{WW}^{\hat{4}}(y)Y_{ZZ}^{\hat{4}}(y)
 \frac{(\vep_1\cdot\vep_2)(\vep_3^*\cdot\vep_4^*)}{p_{12}^2} \right. \nonumber\\
 &&\hspace{20mm}\left.
 +Y_{WZ}(y)\cdot Y_{WZ}(y)\brc{
 \frac{(\vep_1\cdot\vep_3^*)(\vep_2\cdot\vep_4^*)}{p_{13}^2} 
 +\frac{(\vep_1\cdot\vep_4^*)(\vep_2\cdot\vep_3^*)}{p_{14}^2}} \right], 
 \label{cA:WWZZ}
\eea
where 
\bea
 U_{WZ} \defa \frac{1}{2}\brkt{
 2u_W^{\pm_{\rm L}}u_Z^{3_{\rm L}}+u_W^{\hat{\pm}}u_Z^{\hat{3}},
 2u_W^{\pm_{\rm R}}u_Z^{3_{\rm R}}+u_W^{\hat{\pm}}u_Z^{\hat{3}},
 u_W^{\hat{\pm}}\brkt{u_Z^{3_{\rm L}}+u_Z^{3_{\rm R}}}
 +\brkt{u_W^{\pm_{\rm L}}+u_W^{\pm_{\rm R}}}u_Z^{\hat{3}}}, \nonumber\\
 Y_{WW}^{\hat{4}} \defa e^{-2\sgm}
 \brc{\brkt{u_W^{\pm_{\rm L}}-u_W^{\pm_{\rm R}}}'u_W^{\hat{\pm}}
 -\brkt{u_W^{\pm_{\rm L}}-u_W^{\pm_{\rm R}}}\brkt{u_W^{\hat{\pm}}}'}, 
 \nonumber\\
 Y_{ZZ}^{\hat{4}} \defa e^{-2\sgm}
 \brc{\brkt{u_Z^{3_{\rm L}}-u_Z^{3_{\rm R}}}'u_Z^{\hat{3}}
 -\brkt{u_Z^{3_{\rm L}}-u_Z^{3_{\rm R}}}\brkt{u_Z^{\hat{3}}}'}, 
 \nonumber\\
 Y_{WZ} \defa \frac{e^{-2\sgm}}{2}\left(
 2\brkt{u_W^{\pm_{\rm L}}}'u_Z^{3_{\rm L}}+\brkt{u_W^{\hat{\pm}}}'u_Z^{\hat{3}}
 -2u_W^{\pm_{\rm L}}\brkt{u_Z^{3_{\rm L}}}'-u_W^{\hat{\pm}}\brkt{u_Z^{\hat{3}}}',
 \right. \nonumber\\
 &&\hspace{10mm}
 2\brkt{u_W^{\pm_{\rm R}}}'u_Z^{3_{\rm R}}+\brkt{u_W^{\hat{\pm}}}'u_Z^{\hat{3}}
 -2u_W^{\pm_{\rm R}}\brkt{u_Z^{3_{\rm R}}}'-u_W^{\hat{\pm}}\brkt{u_Z^{\hat{3}}}',
 \nonumber\\ 
 &&\hspace{10mm} 
 \brkt{u_W^{\pm_{\rm L}}+u_W^{\pm_{\rm R}}}'u_Z^{\hat{3}}
 +\brkt{u_W^{\hat{\pm}}}'\brkt{u_Z^{3_{\rm L}}+u_Z^{3_{\rm R}}} \nonumber\\
 &&\hspace{20mm} \left.
 -\brkt{u_W^{\pm_{\rm L}}+u_W^{\pm_{\rm R}}}\brkt{u_Z^{\hat{3}}}'
 -u_W^{\hat{\pm}}\brkt{u_Z^{3_{\rm L}}+u_Z^{3_{\rm R}}}' \right). 
\eea

\begin{figure}[t]
\centering  \leavevmode
\includegraphics[width=70mm]{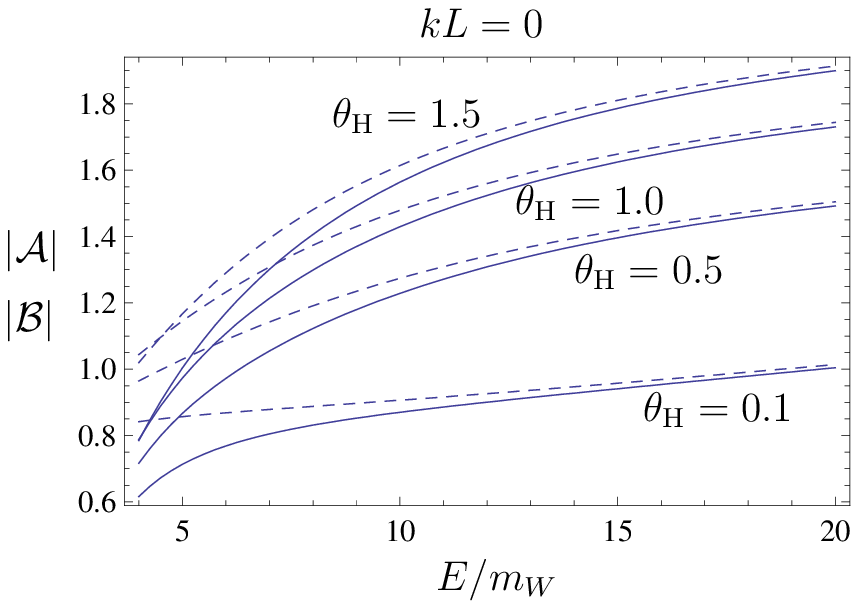} \hspace{10mm}
\includegraphics[width=70mm]{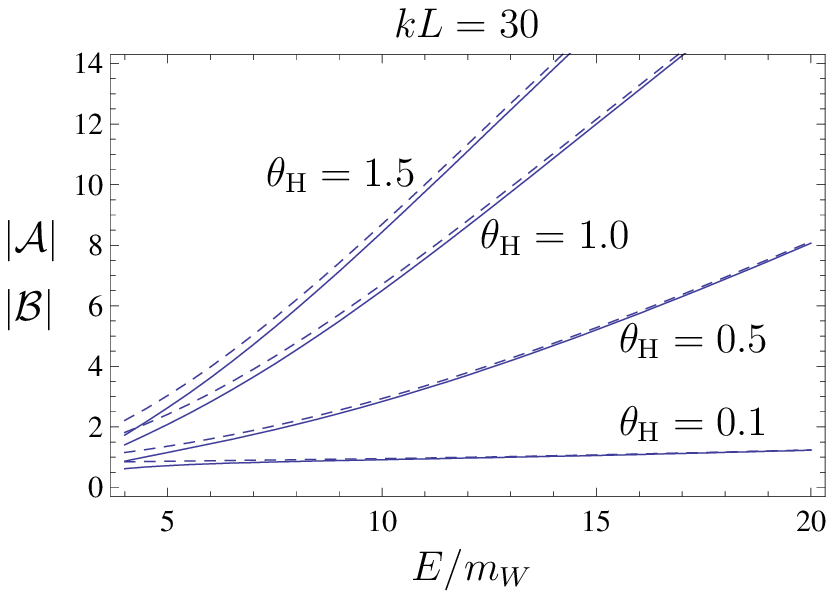}
\caption{The energy dependence of the amplitudes for 
$W_L^++W_L^-\to Z_L+Z_L$. 
The solid lines represent the vector mode scattering~$\cA$, 
and the dashed lines are the gauge-scalar mode scattering~$\cB$. 
The scattering angle is chosen as $\chi=\pi/3$. 
}
\label{amp_WWZZ1}
\end{figure}

Fig.~\ref{amp_WWZZ1} shows the energy dependence of the scattering amplitudes 
in the unit of $m_W$. 
The solid and dashed lines represent the amplitudes 
for the vector modes~$\cA$ and for the gauge-scalar modes~$\cB$, respectively. 
We can explicitly see that the equivalence theorem holds both 
in the flat and warped cases, and $\abs{\cB}-\abs{\cA}=\cO(m_W^2/E^2)$. 
In the warped case ($kL=30$), the situation is similar to 
the $SU(3)$ toy model~\cite{HSY}. 
The amplitudes behave as $E^2$ and grow faster for larger values of $\sin^2\thH$. 
In the flat case ($kL=0$), on the other hand, the situation is quite 
different from the $SU(3)$ model. 
In contrast to the $SU(3)$ model, the amplitudes monotonically increase 
and depend on $\thH$. 
Again, they grow faster for larger values of $\sin^2\thH$. 
This difference from the $SU(3)$ model stems from the mixing between 
different KK levels mentioned around Eq.(\ref{L_mass}). 
Note that $\thH=\cO(1)$ is experimentally excluded in the flat spacetime 
because it leads to too light KK excitation modes. 
However we will also plot the amplitudes for such values of $\thH$ 
in the following, in order to understand theoretical structure 
of the gauge-Higgs unification model. 

These behaviors of the amplitudes reflect the $\thH$-dependences of 
the coupling constants among the gauge and Higgs modes and of $\mKK$. 
First of all, we should notice that the model reduces to SM 
when $\thH\ll 1$ irrespective of the 5D geometry. 
Every coupling constant in the gauge-Higgs sector takes almost the SM value 
and the KK modes are heavy enough to decouple. 
Thus the amplitude takes the same value as SM. 
Namely the amplitudes are almost constant for $E^2\gg m_W^2$. 
When $\thH=\cO(1)$, the coupling constants deviate 
from the SM values~\cite{GHU:HS1,GHU:HS2}. 
In the flat spacetime, the $WWZ$ and $WWZZ$ couplings become smaller  
while the $WWH$ and $ZZH$ couplings take the SM values.  
In the warped spacetime, the latter couplings are suppressed by 
a factor~$\cos\thH$ while the former couplings are almost unchanged 
from the SM values. 
Therefore the $\cO(E^2)$ contributions miss to be cancelled 
among the low-lying modes, and the amplitudes grows. 
For larger $\sin^2\thH$, the deviation of the couplings become larger, 
and thus the amplitudes grow faster. 

\begin{figure}[t]
\centering  \leavevmode
\includegraphics[width=70mm]{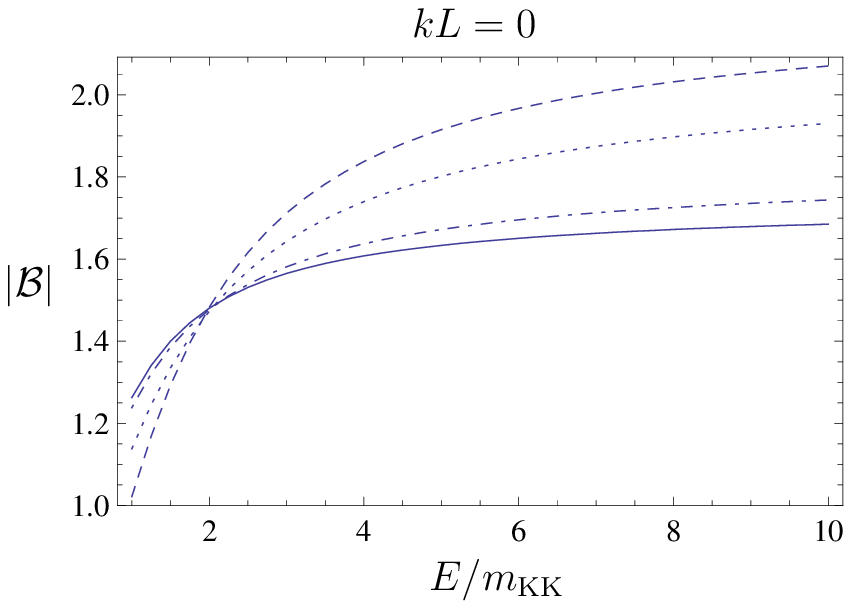} \hspace{10mm}
\includegraphics[width=70mm]{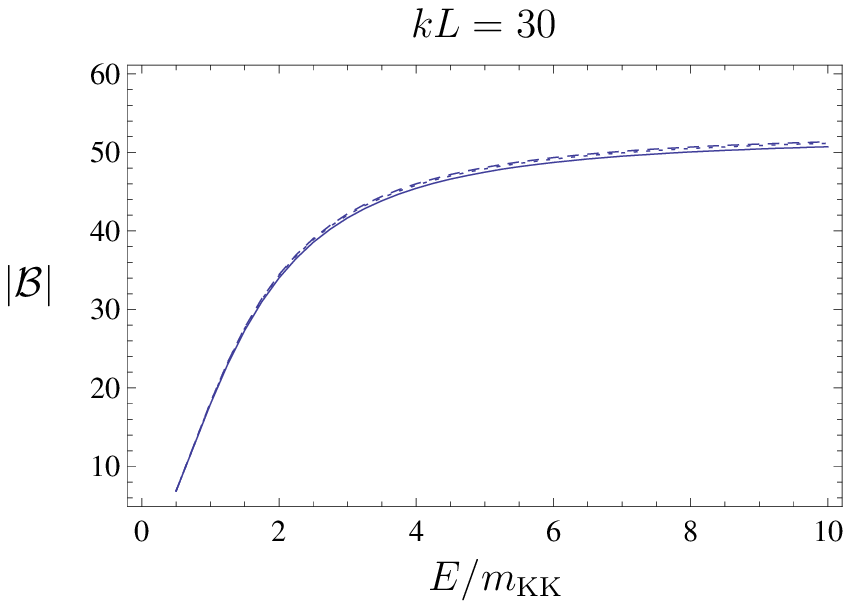}
\caption{The energy dependence of the amplitude~$\cB_{WWZZ}$ 
in the unit of $\mKK$. 
The solid, dotdashed, dotted and dashed lines correspond to 
$\thH=0.1,0.5,1.0,1.5$, respectively. 
}
\label{amp_WWZZ2}
\end{figure}

The remaining $\cO(E^2)$ contribution is eventually cancelled 
by contributions from the KK modes. 
Namely, the amplitudes cease to increase and approach to constant values 
when the KK modes start to propagate. 
We can see this behavior by rescaling the unit of the horizontal axes 
in Fig.~\ref{amp_WWZZ1} to $\mKK$ (Fig.~\ref{amp_WWZZ2}). 
In the warped case, the $\thH$-dependence almost disappears 
in Fig.~\ref{amp_WWZZ2}. 
The $\thH$-dependence appearing in Fig.~\ref{amp_WWZZ1} is cancelled by 
that of $\mKK$. 
Thus the asymptotic constant value of the amplitude is almost determined 
only by the value of $kL$. (See Fig.~2 in Ref.~\cite{HSY}.)

All the above behaviors can also be seen in other processes, such as 
the elastic scatterings:~$W_L^++W_L^-\to W_L^++W_L^-$ and 
$W_L^++Z_L\to W_L^++Z_L$. 
In contrast to the process:~$W_L^++W_L^-\to Z_L+Z_L$, 
there are $s$-channel diagrams exchanging the KK vector bosons 
in these processes, which lead to the resonances. 
The tree-level amplitudes diverges there. 
In order to evaluate the amplitudes around the resonances, 
we have to include the widths of each states, which are obtained 
from one-loop correction of the 5D propagators.

\subsubsection{Forward scattering} \label{fwd_scat}
Next we consider the forward (backward) scattering, 
\ie, $\chi\simeq 0\;(\pi)$. 
Let us first consider the inelastic scattering process:~$W_L^++W_L^-\to Z_L+Z_L$. 
In this case, an $\cO(E^2)$ contribution remains 
and the amplitude monotonically increases even above $m_{\rm KK}$. 
This is because the power counting of $E$ for the amplitude changes 
around $\chi=0$. 
For example, the brace part of $\cA^{\rm S}$ in Eq.(\ref{cA:WWZZ}) 
is expanded (for nonzero $\sin\chi$) as 
\bea
 A_{tu} \defa 
 \frac{(\vep_1\cdot\vep_3^*)(\vep_2\cdot\vep_4^*)}{p_{13}^2}
 +\frac{(\vep_1\cdot\vep_4^*)(\vep_2\cdot\vep_3^*)}{p_{14}^2} \nonumber\\
 \eql \frac{E^2}{4m_W^2m_Z^2}-\frac{m_W^2+m_Z^2}{2m_W^2m_Z^2} 
 +\frac{2m_W^2m_Z^2+(m_W^4+m_Z^4)\cos(2\chi)}{m_W^2m_Z^2E^2\sin^2\chi}
 +\cO(E^{-4}). 
\eea
This means that the expansion becomes invalid when $\sin\chi\simlt\cO(m_W/E)$. 
At $\chi=0$, this quantity reduces to 
\bea
 A_{tu} \eql \frac{(m_W^4+m_Z^4)E^2}{2m_W^2m_Z^2(m_Z^2-m_W^2)^2}
 -\frac{2(m_W^2+m_Z^2)}{(m_Z^2-m_W^2)^2}, 
\eea
and the leading term for the high energy expansion changes. 
Therefore an $\cO(E^2)$ contribution is left in the total amplitude. 
Similar behavior of the amplitude is observed also in SM. 

Next we consider the elastic scattering process:~$W_L^++W_L^-\to W_L^++W_L^-$. 
The mode functions in Eqs.(\ref{cA^C})-(\ref{cA^S}) are taken as 
\bea
 u_m \eql u_k = 
 (u_W^{+_{\rm L}},u_W^{+_{\rm R}},u_W^{\hat{+}},0,0,0,0,0,0,0,0), 
 \nonumber\\
 u_n \eql u_l = 
 (0,0,0,u_W^{-_{\rm L}},u_W^{-_{\rm R}},u_W^{\hat{-}},0,0,0,0,0). 
\eea
Then the expression of the amplitude is reduced to 
\bea
 \cA^{\rm C}_{WWWW} \eql -ig_A^2\int_0^L\dr y\;U_{WW}^2(y)
 \brc{(\vep_1\cdot\vep_2)(\vep_3^*\cdot\vep_4^*)
 +(\vep_1\cdot\vep_3^*)(\vep_2\cdot\vep_4^*)
 -2(\vep_1\cdot\vep_4^*)(\vep_2\cdot\vep_3^*)}, \nonumber\\
 \cA^{\rm V}_{WWWW} \eql -ig_A^2\int_0^L\dr y\int_0^L\dr y'\;
 U_{WW}(y)\cdot G_{\rm T}^{\rm nt}(y,y',\abs{p_{12}})\cdot U_{WW}(y)P_{1234}
 \nonumber\\
 &&+ig_A^2\int_0^L\dr y\int_0^L\dr y'\;
 U_{WW}(y)\cdot G_{\rm T}^{\rm nt}(y,y',\abs{p_{13}})\cdot U_{WW}(y)P_{1324}, 
 \nonumber\\
 \cA^{\rm S}_{WWWW} \eql ig_A^2\int_0^L\dr y\;e^{2\sgm(y)}
 \brkt{Y_{WW}^{\hat{4}}(y)}^2
 \brc{\frac{(\vep_1\cdot\vep_2)(\vep_3^*\cdot\vep_4^*)}{p_{12}^2} 
 +\frac{(\vep_1\cdot\vep_3^*)(\vep_2\cdot\vep_4^*)}{p_{13}^2}}, 
 \label{cA:WWWW}
\eea
where 
\bea
 U_{WW} \defa 
 \frac{1}{2}\brkt{2\brkt{u_W^{\pm_{\rm L}}}^2+\brkt{u_W^{\hat{\pm}}}^2,
 2\brkt{u_W^{\pm_{\rm R}}}^2+\brkt{u_W^{\hat{\pm}}}^2,0,
 2\brkt{u_W^{\pm_{\rm L}}+u_W^{\pm_{\rm R}}}u_W^{\hat{\pm}}}. 
\eea
In this case, the amplitude~$\cA_{WWWW}(E,\chi)$ has a singularity 
at $\chi=0$. 
This is due to the $t$-channel diagram exchanging the massless photon, 
which is proportional to $1/p_{13}^2=\brc{(E^2/2-2m_W^2)(1-\cos\chi)}^{-1}$. 
In any actual collider experiments, however, 
such forward scattering processes cannot be 
measured because they cannot be distinguished from the ones that 
two particles pass by without interacting with each other. 
They are also irrelevant in the cosmological processes by the same reason. 
Therefore the divergence at $\chi=0$ does not lead to any difficulties 
in most practical calculations. 
However we have to deal with this singularity in a proper manner 
when we estimate the unitarity bound. 
We will come back to this point in the next section.

\section{Unitarity bound} \label{unit_bound}
\subsection{Unitarity conditions} \label{untr_cond}
The unitarity bound originates from the unitarity condition of 
the $S$ matrix, $\cS^\dagger\cS=1$, which, with the definition of $\cS=1+i\cT$, 
can be expressed as $\cT^\dagger\cT=2\Im\cT$. 
Taking the matrix element of both sides of the latter relation 
between identical 2-body states and inserting a complete set of 
intermediate states into the left-hand side, we obtain 
\be
 \int_{{\rm PS}_2}\abs{\cT_{\rm el}[2\to 2]}^2
 +\sum_N\int_{{\rm PS}_N}\abs{\cT_{\rm inel}[2\to N]}^2 
 = 2\Im \cT_{\rm el}[2\to 2],  \label{unitarity_cond1}
\ee
where $\cT_{\rm el}[2\to 2]$ and $\cT_{\rm inel}[2\to N]$ denote 
amplitudes for a 2-body elastic scattering 
and for an inelastic scattering with $N$-body final state respectively, 
and $\int_{{\rm PS}_N}$ denotes the $N$-body phase space integration. 
The right-hand side is evaluated in the forward direction.  
By performing the partial wave expansion for the scattering amplitudes
for the $2\to 2$ processes, Eq.(\ref{unitarity_cond1}) is rewritten as 
(see, for example, Ref.~\cite{unt_bd})
\bea
 &&\sum_{j=0}^\infty (2j+1)\brc{
 \frac{1}{\rho_{\rm e}}\brkt{\frac{\rho_{\rm e}^2}{4}
 -\abs{\eta_{\rm f}^{\rm el}a_j^{\rm el}-\frac{i\rho_{\rm e}}{2}}^2}
 -\sum_{N=2}\frac{\eta_{\rm f}^{\rm el}\eta_{\rm f}^{\rm inel}}
 {\rho_{\rm i}}\abs{a_j^{\rm inel}}^2} \nonumber\\
 \eql \frac{\eta_{\rm f}^{\rm el}}{32\pi}\sum_{N\neq 2}
 \int_{{\rm PS}_N}\abs{\cT_{\rm inel}[2\to N]}^2 > 0, 
 \label{unitarity_cond2}
\eea
where the symmetry factors~$\rho_{\rm e}$ and $\rho_{\rm i}$ 
equal $1!$ ($2!$) if the 2-body final state consists of nonidentical 
(identical) particles for the elastic and inelastic scattering processes. 
The partial wave components of the amplitudes are defined as\footnote{
Here we focus on the case that the two particles in the initial or final state 
have the same helicity. 
}  
\bea
 a_j^{\rm el} \defa \frac{1}{32\pi}\int_{-1}^1\dr(\cos\chi)P_j(\cos\chi)
 \cT_{\rm el}[2\to 2], \nonumber\\
 a_j^{\rm inel} \defa \frac{1}{32\pi}\int_{-1}^1\dr(\cos\chi)P_j(\cos\chi)
 \cT_{\rm inel}[2\to 2], 
\eea
where $P_j(x)$ are the Legendre polynomials. 
The factors~$\eta_{\rm f}^{\rm el}$ and $\eta_{\rm f}^{\rm inel}$ are 
functions of the total energy and the masses of the final state particles 
defined as 
\be
 \eta(E,m_l,m_k) \equiv \frac{2p_F}{E} = 
 \brc{1-\frac{2(m_l^2+m_k^2)}{E^2}+\frac{(m_l^2-m_k)^2}{E^4}}^{1/2}, 
 \label{def:eta}
\ee
evaluated for the elastic and inelastic scattering processes, respectively. 
In the high energy region ($E^2\gg m_l^2,m_k^2$), 
these factors are approximately equal to one. 

In the following we assume that the S-wave component ($j=0$) is dominant 
in Eq.(\ref{unitarity_cond2}). 
Then, for scattering of $W^+$ and $W^-$, 
we obtain the following unitarity condition. 
\bea
 &&\abs{\eta_{WW}^{00}a_0^{00}[WW]-\frac{i}{2}}^2
 +\frac{\eta_{WW}^{00}\eta_{ZZ}^{00}}{2}\abs{a_0^{00}[ZZ]}^2 \nonumber\\
 &&+\eta_{WW}^{00}\sum_{(l,k)\neq(0,0)}
 \brc{\eta_{WW}^{lk}\abs{a_0^{lk}[WW]}^2
 +\frac{\eta_{ZZ}^{lk}}{\rho_{lk}}\abs{a_0^{lk}[ZZ]}^2}
 < \frac{1}{4},  \label{untr_bd1}
\eea
where $a_0^{lk}[WW]$ and $a_0^{lk}[ZZ]$ are the S-wave amplitudes 
for the processes to $W^{+(l)}, W^{-(k)}$ and 
$Z^{(l)}, Z^{(k)}$ in the final state respectively, 
and $\eta_{WW}^{lk}$ and $\eta_{ZZ}^{lk}$ are the corresponding factors 
defined in Eq.(\ref{def:eta}). 
Here $W^{\pm(l)}$ and $Z^{(l)}$ denote the $l$-th KK excitation modes in 
the charged and neutral sectors.\footnote{
In this notation, $Z^{(l)}$ ($l=0,1,2,\cdots$) include 
the KK modes of the photon except for the massless photon. 
The lowest mode~$Z^{(0)}$ is identified with the $Z$ boson. 
} 
The symmetry factor~$\rho_{lk}$ equals $1!$ ($2!$) 
when $l\neq k$ ($l=k$). 
We do not consider processes to fermions in the final state 
because we have not specified the matter sector. 

Here we comment on contributions of the forward scattering 
to the S-wave amplitudes. 
Let us first consider the process:~$W^+_L+W^-_L\to Z_L+Z_L$. 
Since $i\cT_{\rm el}[2\to 2]=\cA_{WWZZ}$ at tree level, 
the S-wave amplitude is obtained as 
\be
 a_0^{00}[ZZ](E) = \frac{-i}{32\pi}\int_{-1}^1\dr (\cos\chi)\;
 \cA_{WWZZ}(E,\chi) 
 = \frac{-i}{16\pi}\int_0^1\dr (\cos\chi)\;\cA_{WWZZ}(E,\chi).  
 \label{a0ZZ}
\ee
In the last equality, we have used the relation 
$\cA_{WWZZ}(E,\chi)=\cA_{WWZZ}(E,\pi-\chi)$. 
As mentioned in Sec.~\ref{fwd_scat}, the integrand grows as $E^2$ 
in the region $1-\abs{\cos\chi}\simlt\cO(m_W^2/E^2)$ 
while it approaches to a constant for $E^2\gg\mKK^2$ in the other region of 
$\cos\chi$. 
Therefore, $a_0^{00}[ZZ]$ behaves as $\cO(E^0)$ at high energies. 
In fact, it grows logarithmically above $\mKK$. 
(See Fig.~3 in Ref.~\cite{HSY}.)

Next we consider the elastic scattering:~$W^+_L+W^-_L\to W^+_L+W^-_L$. 
As mentioned at the end of the previous section, 
the tree-level amplitude~$\cA_{WWWW}(E,\chi)$ diverges at $\chi=0$. 
Such divergence is smeared out by taking into account 
the instability of the W bosons in the final state, 
as shown in Appendix~\ref{fwd_singularity}. 
The effect of the instability is translated into a cut-off 
for the $\cos\chi$-integral. 
Then the S-wave amplitude is calculated as  
\be
 a_0^{00}[WW](E) = \frac{-i}{32\pi}\int_{-1}^{x_{\rm cut}}\dr (\cos\chi)\;
 \cA_{WWWW}(E,\chi),  \label{a0WW}
\ee
where $x_{\rm cut}$ is given by Eq.(\ref{def:x_cut}). 

\begin{figure}[t]
\centering  \leavevmode
\includegraphics[width=70mm]{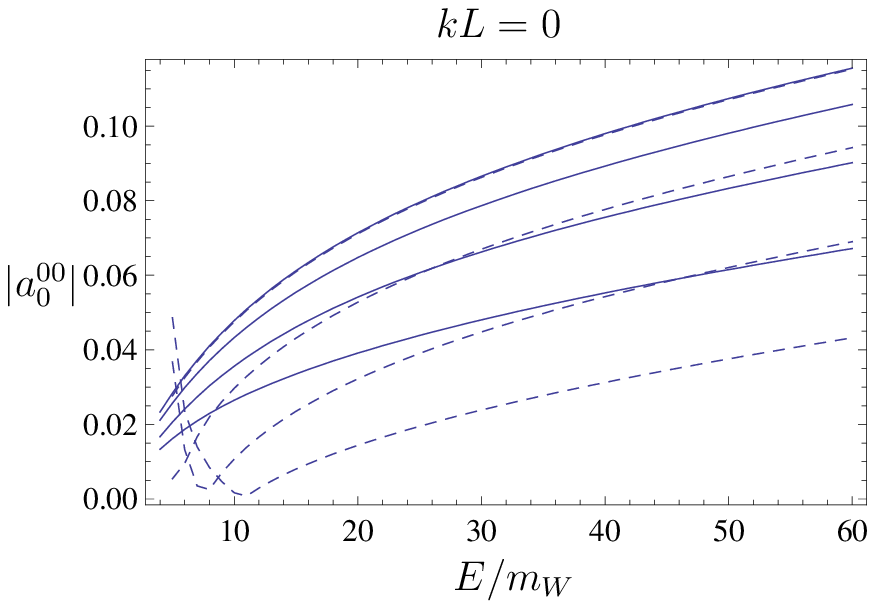} \hspace{10mm}
\includegraphics[width=70mm]{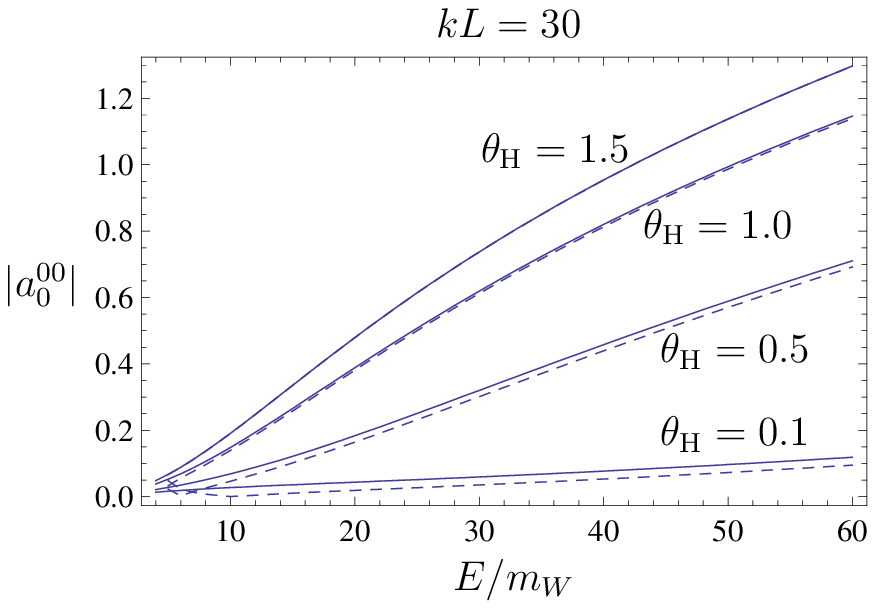}
\caption{The S-wave amplitude for $W_L^++W_L^-\to Z_L+Z_L$. 
The solid (dashed) lines represent 
$a_0^{00}[ZZ]$ with $m_H=2m_W$ ($4m_W$)
for $\thH=0.1,0.5,1.0,1.5$ from bottom to top. 
}
\label{Swave1}
\end{figure}

Notice that the Higgs boson is massless at tree-level 
in the gauge-Higgs unification scenario. 
Thus the $t$-channel diagram exchanging the Higgs boson is also singular 
at $\chi=0$. 
Therefore the Higgs mass has to be incorporated 
in a proper manner in order to evaluate the S-wave amplitude. 
The consistent way to deal with the nonzero Higgs mass is 
to include quantum corrections, which is however 
beyond the scope of this paper. 
Instead, we introduce the Higgs mass parameter~$m_H$  
in the Higgs propagator appearing in the expressions of the amplitude, 
as a free parameter as discussed in the introduction. 
Namely, we modify the Higgs-propagator part of 
$\cA^{\rm S}_{WWZZ}$ in Eq.(\ref{cA:WWZZ}) as 
\bea
 Y_{WW}^{\hat{4}}(y)Y_{ZZ}^{\hat{4}}(y)
 \frac{(\vep_1\cdot\vep_2)(\vep_3^*\cdot\vep_4^*)}{p_{12}^2} 
 \toa Y_{WW}^{\hat{4}}(y)Y_{ZZ}^{\hat{4}}(y)
 \frac{(\vep_1\cdot\vep_2)(\vep_3^*\cdot\vep_4^*)}{p_{12}^2+m_H^2}, 
\eea 
and of $\cA^{\rm S}_{WWWW}$ in Eq.(\ref{cA:WWWW}) as 
\bea
 &&\brkt{Y_{WW}^{\hat{4}}(y)}^2\brc{\frac{(\vep_1\cdot\vep_2)
 (\vep_3^*\cdot\vep_4^*)}{p_{12}^2}+\frac{(\vep_1\cdot\vep_3^*)
 (\vep_2\cdot\vep_4^*)}{p_{13}^2}} \nonumber\\
 \toa 
 \brkt{Y_{WW}^{\hat{4}}(y)}^2\brc{\frac{(\vep_1\cdot\vep_2)
 (\vep_3^*\cdot\vep_4^*)}{p_{12}^2+m_H^2}+\frac{(\vep_1\cdot\vep_3^*)
 (\vep_2\cdot\vep_4^*)}{p_{13}^2+m_H^2}}. 
\eea
This is a good approximation since the quantum corrections 
to the KK masses are subdominant and thus negligible. 
Fig.~\ref{Swave1} shows the $m_H$-dependence of 
the S-wave amplitude~$a_0^{00}[ZZ](E)$. 
We can see from these figures that the $m_H$-dependence disappears 
when $\thH=\pi/2$ in both the flat and warped cases. 
This is because the $WWH$ and $ZZH$ couplings vanish and 
the Higgs propagator does not contribute to the amplitude when $\thH=\pi/2$. 
For other values of $\thH$, 
the introduction of larger $m_H$ reduces the amplitude.

\subsection{Unitarity bound from WW scattering}
Now we estimate the unitarity bound. 
Let us define the summed amplitude~$\bar{a}_0(E)$ as  
\bea
 \bar{a}_0 \defa \brc{\brkt{\eta_{WW}^{00}\Re a_0^{00}[WW]}^2
 +\frac{\eta_{WW}^{00}\eta_{ZZ}^{00}}{2}\abs{a_0^{00}[ZZ]}^2}^{1/2}. 
 \label{def:bar_a0}
\eea
Then the following unitarity bound is obtained from Eq.(\ref{untr_bd1}). 
\be
 \bar{a}_0(E) < \frac{1}{2}.  \label{untr_bd2}
\ee
Notice that the left-hand side of Eq.(\ref{untr_bd1}) 
already saturates the unitarity bound if $\Im a_0^{00}[WW]=0$. 
Although the imaginary part of the S-wave amplitudes are zero 
at tree level,\footnote{
To be precise, there is a small contribution to $\Im a_0^{00}[WW]$ 
coming from the principal value integral~(\ref{PVInt}) 
in Appendix~\ref{fwd_singularity}. }
nonvanishing contribution comes out at loop level. 
This loop contribution can be large near $\Lmd_{\rm uni}$ 
since perturbative expansion is less reliable there. 
Hence we should take it into account in order to obtain 
a nontrivial unitarity bound~\cite{DJL}. 
However estimation of $\Im a_0^{00}[WW]$ at loop level is beyond 
the scope of this paper. 
In this paper, we simply assume that there is enough contribution 
to $\Im a_0^{00}[WW]$ at loop level to cancel $-i/2$ in the first term 
of the left-hand side of Eq.(\ref{untr_bd1}), 
and consider only the real part of the S-wave amplitudes 
to estimate the unitarity bound. 

\begin{figure}[t]
\centering  \leavevmode
\includegraphics[width=70mm]{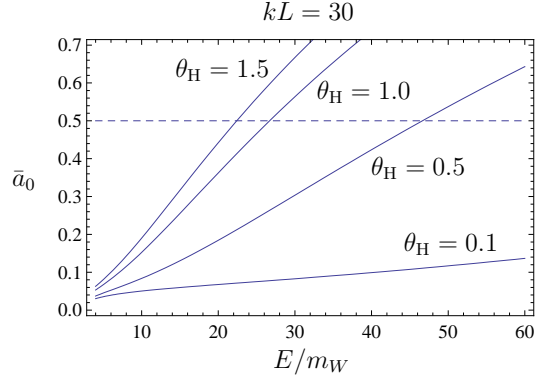}
\caption{The summed amplitude~$\bar{a}_0$ defined in Eq.(\ref{def:bar_a0}) 
in the warped spacetime. 
The Higgs mass is chosen as $m_H=2m_W$. 
The dashed line represents the unitarity bound. 
}
\label{Swave_warp2}
\end{figure}

Fig.~\ref{Swave_warp2} shows $\bar{a}_0(E)$ for various values of $\thH$ 
in the warped spacetime. 
The Higgs mass is chosen as $m_H=2m_W$ in this plot. 
The dashed line represents the unitarity bound. 
From this figure, we can read off the (conservative) unitarity violation 
scale as 
$\Lmd_{\rm uni}\simeq 22m_W\simeq 1.8$~TeV for $\thH=1.5$, and 
$\Lmd_{\rm uni}\simeq 46m_W\simeq 3.7$~TeV for $\thH=0.5$. 
The unitarity is violated at $\cO(1~\mbox{TeV})$ 
when $\thH=\pi/2$ since the $WWH$ and $ZZH$ couplings vanish. 
We cannot see any KK resonances in Fig.~\ref{Swave_warp2} 
despite the fact that 
the amplitude~$\cA_{WWWW}$ has divergent peaks at the resonances, 
which correspond to the KK gauge bosons. 
The reason for this is as follows. 
Such divergent peaks originate from the $s$-channel diagrams 
corresponding to a term proportional to $P_{1234}$ in Eq.(\ref{cA:WWWW}). 
However this term will vanish after integrating for $\cos\chi$ 
over $[-1,1]$ because it is proportional to $\cos\chi$.\footnote{
The cut-off~$x_{\rm cut}$ in Eq.(\ref{a0WW}) is introduced 
only the integral of the $t$-channel contribution, which corresponds to 
the term proportional to $P_{1324}$ in Eq.(\ref{cA:WWWW}). 
}
This fact can also be understood from the viewpoint of 
the spin composition. 
Since the longitudinal vector boson is a state with the angular 
momentum~$(j,j_3)=(1,0)$, intermediate KK vector boson states 
for the $s$-channel must also have the quantum number~$(j,j_3)=(1,0)$. 
When the orbital angular momentum is zero, however, 
it is impossible to creat such a spin state 
by the composition of two states with $(j,j_3)=(1,0)$. 
Therefore, the $s$-channel contribution to the S-wave amplitude is zero. 

\begin{figure}[t]
\centering  \leavevmode
\includegraphics[width=70mm]{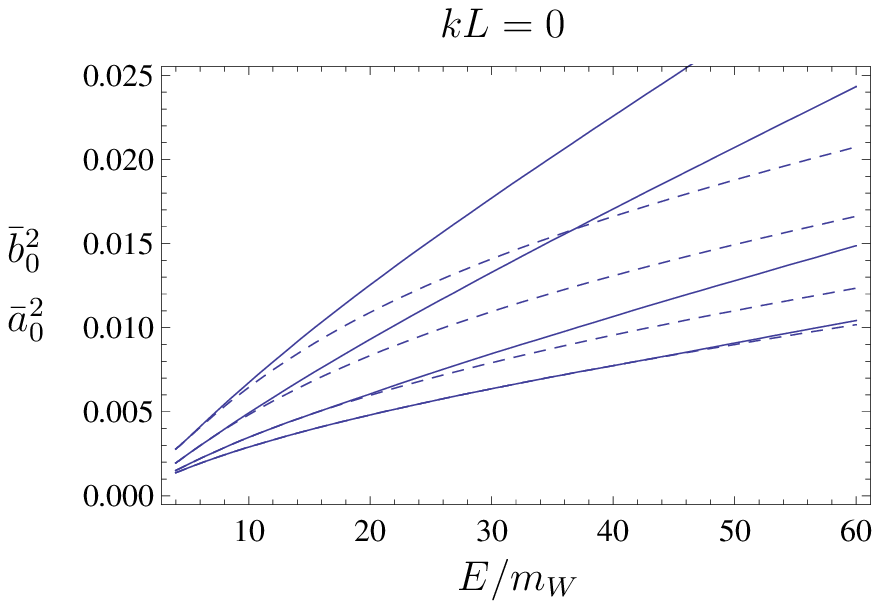} \hspace{10mm}
\includegraphics[width=70mm]{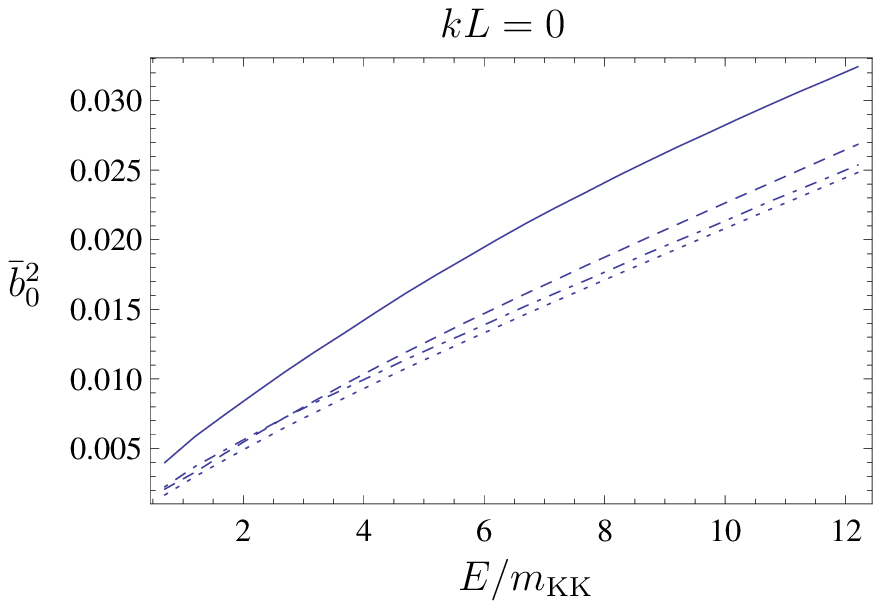}
\caption{The summed amplitudes in the flat spacetime. 
In the left figure, the solid lines denote $\bar{b}_0^2$ 
defined in Eq.(\ref{def:bar_b0}), and 
the dashed lines denote $\bar{a}_0^2$ defined in Eq.(\ref{def:bar_a0}),  
for $\thH=0.2,0.5,1.0,1.5$ from bottom to top. 
In the right figure, the solid, dotdashed, dotted and dashed lines 
correspond to $\thH=0.2,0.5,1.0,1.5$, respectively. 
The Higgs mass is chosen as $m_H=2m_W$ in both figures. 
}
\label{Swave_flat2}
\end{figure}

In the flat spacetime, the amplitude grows slowly and 
thus $\Lmd_{\rm uni}$ is much higher than the warped case. 
In fact, the unitarity bound from Eq.(\ref{untr_bd2}) is 
determined by the logarithmic behavior of $\bar{a}_0(E)$ at high energies, 
which is mentioned below Eq.(\ref{a0ZZ}). 
In such a case, contributions of inelastic scattering 
involving the KK modes in the final state become important 
because a large number of scattering processes are kinematically allowed 
near $\Lmd_{\rm uni}$. 
Therefore the summed amplitude~$\bar{a}_0$ should be modified 
by including the contributions of such processes as 
\bea
 \bar{b}_0^2 \defa \brkt{\eta_{WW}^{00}\Re a_0^{00}[WW]}^2
 +\frac{\eta_{WW}^{00}\eta_{ZZ}^{00}}{2}\abs{a_0^{00}[ZZ]}^2 \nonumber\\
 &&+\eta_{WW}^{00}\sum_{l\geq 1}\brkt{
 \eta_{WW}^{ll}\abs{a_0^{ll}[WW]}^2
 +\frac{\eta_{ZZ}^{ll}}{2}\abs{a_0^{ll}[ZZ]}^2}. 
 \label{def:bar_b0}
\eea
Contributions of the scattering processes to different 
KK levels~$a_0^{lk}[WW]$ and $a_0^{lk}[ZZ]$ ($l\neq k$) are generically small 
and can be neglected. 
The unitarity bound is written as 
\be
 \bar{b}_0^2(E) < \frac{1}{4}. 
\ee
The left plot in Fig.~\ref{Swave_flat2} shows 
the energy dependence of $\bar{b}_0$ (solid lines) 
and $\bar{a}_0$ (dashed lines). 
We can see from this plot that the summed amplitude~$\bar{b}_0^2(E)$ 
asymptotically behaves as an increasing linear function, 
while $\bar{a}_0^2(E)$ does logarithmically. 
This is a consequence of the intrinsic nonrenormalizability of 
the higher dimensional gauge theory, 
as was pointed out in Ref.~\cite{NDA} in the context of 
the Higgsless models. 
The inclination of the asymptotic line vary over the values of $\thH$.  
It is mainly determined by the KK mass scale~$\mKK$. 
For smaller values of $\thH$, the KK modes does not appear 
until higher energy scales, and the amplitude grows at a slower pace. 
This can be explicitly seen in the right plot of Fig.~\ref{Swave_flat2}, 
in which the unit of the horizontal axis is rescaled to $\mKK$. 
In this plot, the inclination of the asymptotic line is almost 
independent of $\thH$. 
By extrapolating the asymptotic lines, 
we obtain the unitarity violation scale as 
$\Lmd_{\rm uni}\simeq 140\mKK$, irrespective of the value of $\thH$. 

In Ref.~\cite{NDA}, it was found that $\Lmd_{\rm uni}$ is 
roughly equal (up to a small numerical factor) 
to the cut-off scale of the 5D theory obtained 
from naive dimensional analysis (NDA) in the Higgsless model. 
In our model, the NDA cut-off scale~$\Lmd_{\rm NDA}$ is estimated as 
\be
 \Lmd_{\rm NDA} = \frac{24\pi^3}{g^2L} \simeq 592\mKK 
 \simeq \frac{1.86\times 10^3 m_W}{\sin^{-1}\brkt{\frac{1}{\sqrt{2}}\sin\thH}}. 
\ee
Here we have used that 
$g^2=g_A^2/L=0.4$ is the 4D weak gauge coupling constant, 
and Eq.(\ref{mWZ:flat}). 
Therefore $\Lmd_{\rm uni}$ is lower than $\Lmd_{\rm NDA}$ 
by about a factor of four in our model. 

Finally we remark that $\Lmd_{\rm uni}$ estimated here is a conservative one 
since we did not consider scattering processes to fermions in the final states, 
as mentioned in the introduction.

\section{Summary} \label{summary}
We have estimated a scale~$\Lmd_{\rm uni}$, 
at which the tree-level unitarity is violated, 
in the 5D $SO(5)\times\ubl$ gauge-Higgs unification model 
by evaluating amplitudes of the weak boson scattering. 
The 5D propagators are useful to evaluate the amplitudes 
because we need not explicitly calculate the KK mass eigenvalues 
and mode functions nor perform infinite summation over 
the KK modes propagating in the internal lines. 
In particular above the KK mass scale~$\mKK$, they provide a practical method 
of evaluating the amplitudes. 
Although inelastic scattering processes to fermionic final states are not 
examined, the techniques illustrated in this article are also useful to 
evaluate them, and similar behaviors of the amplitudes are expected even 
when they are incorporated, while the numerical value of $\Lmd_{\rm uni}$ 
would be somewhat reduced.

We have numerically checked the equivalence theorem between the amplitudes 
for the 4D longitudinal vector modes 
and for the gauge-scalar modes. 
The amplitude with nonzero scattering angle monotonically increases 
up to $\mKK$, and depends on the Wilson line phase~$\thH$. 
It grows faster for larger values of $\sin^2\thH$. 
In the warped spacetime, its value is enhanced for $\thH=\cO(1)$ 
while it is reduced to that in the flat spacetime for $\thH\ll 1$. 
These behaviors can be understood by the $\thH$-dependences of 
the coupling constants among the gauge and Higgs modes and of $\mKK$. 
The growth of the amplitude stems from deviation of the coupling constants 
from the SM value. 
In the warped case, for example, the $WWH$ and $ZZH$ couplings are 
suppressed from the SM values by a factor of $\cos\thH$. 
Due to this deviation, the $\cO(E^2)$ contributions of $\cA^{\rm C}$, 
$\cA^{\rm V}$ and $\cA^{\rm S}$ in Eqs.(\ref{cA^C})-(\ref{cA^S}) 
miss to be cancelled among 
the light modes, and the amplitude grows as shown in Fig.~\ref{amp_WWZZ1}.  
The remaining $\cO(E^2)$ contribution depends on the deviation of 
the couplings and has the maximal value when $\thH=\pi/2$. 
It is eventually cancelled by the contributions from the KK gauge bosons. 
Then the amplitude ceases to increase and approaches to a constant value 
above $\mKK$. 
(See Fig.~\ref{amp_WWZZ2}). 
These behaviors are also observed in the $SU(3)$ model~\cite{HSY}, 
and are thought to be common to the gauge-Higgs unification models.  
In contrast to the $SU(3)$ model, however, the amplitude in our model 
grows and depends on $\thH$ even in the flat spacetime. 
This difference originates from the mixing between different KK levels 
mentioned around Eq.(\ref{L_mass}). 

In Ref.~\cite{FPR}, three separate scales that determine the dynamics 
of the scattering processes are introduced, \ie, 
the electroweak breaking scale~$v$, 
the Higgs boson decay constant~$f_h$,\footnote{
This is the composite scale of the Higgs boson in the holographic dual picture.} 
and the KK mass scale~$\mKK$. 
In our notation, these scales are related to each other as 
$v=f_h\thH$ and $f_h=\sqrt{2}/(g_A\sqrt{L})=\sqrt{2}\mKK/(\pi g)$ 
in the flat case, and 
$v=f_h\sin\thH$ and $f_h\simeq 2\sqrt{k}e^{-kL}/g_A 
\simeq 2\mKK/(\pi g\sqrt{kL})$ in the warped case.\footnote{
These correspondence are essentially the same as in the $SU(3)$ model 
in Ref.~\cite{HSY}. }
In the terminology of Ref.~\cite{FPR}, the case of $\thH\ll 1$ is referred to 
as the `Higgs limit', and the case of $\thH\simeq \pi/2$ is as 
the `Higgsless limit'. 
The Higgs boson mainly unitarizes the scattering processes in the former 
while it does not in the latter. 

We have evaluated the S-wave amplitudes in order to estimate $\Lmd_{\rm uni}$. 
We considered the scattering of $W^+_L$ and $W^-_L$,\footnote{
The unitarity bound from other processes, 
such as the scattering of $W^\pm_L$ and $Z_L$ or of 
two Z bosons, is weaker. 
}
including possible inelastic scatterings. 
In order to evaluate the S-wave amplitude for the elastic scattering, 
we have to deal with the singularity of the amplitude at forward scattering 
in a proper manner. 
We have accomplished this 
by taking into account the instability of the W bosons in the final state. 
The results are depicted in Figs.~\ref{Swave_warp2} and \ref{Swave_flat2}. 
From Fig.~\ref{Swave_warp2}, we can read off 
$\Lmd_{\rm uni}\simeq 1.3\mKK\simeq 7f_h$. 
The unitarity is violated at $\cO(1~\mbox{TeV})$ 
for $\thH=\pi/2$ in the warped case because the $WWH$ and $ZZH$ couplings 
vanish and the situation becomes similar 
to SM without the Higgs boson in such a case. 
For $\thH=\cO(0.1)$, the unitarity is maintained up to $\cO(20~\mbox{TeV})$. 
In the flat spacetime, $\Lmd_{\rm uni}$ becomes 
much higher than the warped case. 
In this case, a large number of inelastic scatterings to the KK modes become 
kinematically allowed around $\Lmd_{\rm uni}$, 
and thus we should take into account contributions 
from those scattering processes. 
The summed amplitude~$\bar{b}_0^2(E)$ defined in Eq.(\ref{def:bar_b0}) 
is approximately a linear function in the high energy region. 
(See Fig.~\ref{Swave_flat2}.)
This is a consequence of the intrinsic nonrenormalizability of 
the higher dimensional gauge theory. 
We have found that $\Lmd_{\rm uni}\simeq 140\mKK$, 
which is lower than the cut-off scale~$\Lmd_{\rm NDA}$ 
from naive dimensional analysis by about a factor of four.

\vskip 0.5cm

\leftline{\bf Acknowledgments} \nopagebreak
The authors would like to thank M.~Kurachi and K.~Tobe 
for useful information and discussion. 
This work was supported in part 
by the scientific grant from the ministry of education, 
science, sports, and culture of Japan 
(Nos.20540272, 20039006, 20025004, and 20244028), 
by Special Postdoctoral Researchers Program at RIKEN (Y.S.), 
and the Japan Society for the Promotion of Science (T.Y.).

\appendix

\section{Bases of mode functions} \label{basis_fcn}
Here we define bases of mode functions, 
following Ref.~\cite{Falkowski}. 
The functions~$C_0(y,m)$ and $S_0(y,m)$ are defined as 
two independent solutions to 
\be
 \brkt{\frac{d}{dy}e^{-2\sgm}\frac{d}{dy}+m^2}f=0, 
\ee
with initial conditions 
\bea
 C_0(0,m) \eql 1, \;\;\;
 C'_0(0,m) = 0, \nonumber\\
 S_0(0,m) \eql 0, \;\;\;
 S'_0(0,m) = me^{-\sgm(L)}.  \label{basis_fcn1}
\eea
The prime denotes derivative in terms of $y$. 

For the derivation of 5D propagators in Appendix~\ref{5Dpropagator}, 
it is convenient to define another basis functions~$C_L(y,m)$ 
and $S_L(y,m)$ with initial conditions 
\bea
 C_L(L,m) \eql 1, \;\;\;
 C'_L(L,m) = 0, \nonumber\\
 S_L(L,m) \eql 0, \;\;\;
 S'_L(L,m) = me^{\sgm(L)}. \label{basis_fcn2}
\eea

From the Wronskian relation, the above functions satisfy 
\bea
 && C_0(y,m)S'_0(y,m)-S_0(y,m)C'_0(y,m) \nonumber\\
 \eql C_L(y,m)S'_L(y,m)-S_L(y,m)C'_L(y,m)
 = me^{2\sgm(y)-\sgm(L)}.  \label{Wronskian}
\eea

The two bases are related to each other by 
\bea
 C_L(y,m) \eql \frac{e^{-\sgm(L)}}{m}\brc{S'_0(L,m)C_0(y,m)
 -C'_0(L,m)S_0(y,m)}, \nonumber\\
 S_L(y,m) \eql -\brc{
 S_0(L,m)C_0(y,m)-C_0(L,m)S_0(y,m)}. 
 \label{rel:basis_fcns}
\eea

\begin{description}
\item[Flat spacetime] \mbox{}\\
In the flat spacetime, \ie, $\sgm(y)=0$, 
the basis functions are reduced to 
\bea 
 C_0(y,m) \eql \cos(my), \;\;\;
 S_0(y,m) = \sin(my), \nonumber\\
 C_L(y,m) \eql \cos\brc{m(y-L)}, \;\;\;
 S_L(y,m) = \sin\brc{m(y-L)}. 
\eea

\item[Randall-Sundrum spacetime] \mbox{}\\
In the Randall-Sundrum spacetime, \ie, $\sgm(y)=ky$, 
the basis functions are written in terms of the Bessel functions as 
\bea
 C_0(y,m) \eql \frac{\pi m}{2k}e^{ky}
 \brc{Y_0\brkt{\frac{m}{k}}J_1\brkt{\frac{m}{k}e^{ky}}
 -J_0\brkt{\frac{m}{k}}Y_1\brkt{\frac{m}{k}e^{ky}}}, \nonumber\\
 S_0(y,m) \eql -\frac{\pi m}{2k}e^{k(y-L)}
 \brc{Y_1\brkt{\frac{m}{k}}J_1\brkt{\frac{m}{k}e^{ky}}
 -J_1\brkt{\frac{m}{k}}Y_1\brkt{\frac{m}{k}e^{ky}}}, \nonumber\\
 C_L(y,m) 
 \eql \frac{\pi m}{2k}e^{ky}
 \brc{Y_0\brkt{\frac{m}{k}e^{kL}}J_1\brkt{\frac{m}{k}e^{ky}}
 -J_0\brkt{\frac{m}{k}e^{kL}}Y_1\brkt{\frac{m}{k}e^{ky}}}, \nonumber\\
 S_L(y,m) 
 \eql -\frac{\pi m}{2k}e^{ky}
 \brc{Y_1\brkt{\frac{m}{k}e^{kL}}J_1\brkt{\frac{m}{k}e^{ky}}
 -J_1\brkt{\frac{m}{k}e^{kL}}Y_1\brkt{\frac{m}{k}e^{ky}}}. \nonumber\\
\eea
\end{description}

\section{Derivation of 5D propagators} \label{5Dpropagator}
Here we derive explicit forms of 5D propagators. 
We take the same strategy as in the appendix of Ref.~\cite{GP}. 
Since the 4D vector part~$A_\mu$ and the gauge-scalar part~$A_y$ are decoupled 
at the quadratic level with our choice of the gauge-fixing function, 
the mixed components of 
the propagator~$\langle 0|TA_\mu^I(p,y)A_y^{\bar{J}}(-p,y')|0\rangle$ vanish. 
In this section, we work in the Scherk-Schwarz basis defined 
by Eqs.(\ref{gauge_trf}) and (\ref{def:Omg}). 

\subsection{Vector propagator}
The 5D propagator~$iG_{\mu\nu}^{I\bar{J}}(p,y,y')\equiv
\langle 0|TA_\mu^I(p,y)A_\nu^{\bar{J}}(-p,y')|0\rangle$ satisfies 
\be
 \sbk{\brc{\der_y^2-2\sgm'\der_y-e^{2\sgm}p^2}
 \dlt_\mu^{\;\;\nu}+e^{2\sgm}\brkt{\frac{1}{\xi}-1}
 p_\mu p^\nu}G^{I\bar{J}}_{\nu\rho}(p,y,y') 
 = e^{2\sgm}\eta_{\mu\rho}\dlt^{I\bar{J}}\dlt(y-y'),  
 \label{eq_propagator}
\ee
with boundary conditions, 
\bea
 &&\der_y G_{\mu\nu}^{I\bar{J}} 
 = \der_y\brkt{s_\phi G_{\mu\nu}^{3_{\rm R}\bar{J}}+c_\phi G_{\mu\nu}^{B\bar{J}}}
 = 0,  \;\;\; (I=\pm_{\rm L},3_{\rm L}) \nonumber\\
 &&G_{\mu\nu}^{I\bar{J}} = c_\phi G_{\mu\nu}^{3_{\rm R}\bar{J}}
 -s_\phi G_{\mu\nu}^{B\bar{J}} = 0, \;\;\;
 (I=\pm_{\rm R},\hat{\pm},\hat{3},\hat{4})
 \label{bdcd_propagator:0}
\eea
at $y=0$, and 
\bea
 &&\brkt{R_\theta}^{IK}\der_yG_{\mu\nu}^{K\bar{J}} = 0, \;\;\;
 (I=\pm_{\rm L},\pm_{\rm R},3_{\rm L},3_{\rm R},B) \nonumber\\
 &&\brkt{R_\theta}^{IK}G_{\mu\nu}^{K\bar{J}} = 0, \;\;\;
 (I=\hat{\pm},\hat{3},\hat{4}) 
 \label{bdcd_propagator:L}
\eea
at $y=L$. 
The indices~$I$ and $\bar{J}$ are defined 
in Eqs.(\ref{index_I}) and (\ref{index_barJ}). 
A constant matrix~$R_\theta$ is a rotation matrix for the indices 
of the adjoint representation corresponding to a transformation by 
$\Omg(L)$ defined in Eq.(\ref{def:Omg}), \ie, 
\be
 \brkt{R_\theta}^{IJ}A_M^J = \sbk{\Omg^{-1}(L)\bdm{A}_M\Omg(L)}^I 
 \equiv 2\tr\brc{T^{\bar{I}}\Omg^{-1}(L)\bdm{A}_M\Omg(L)}. 
 \label{def:R_tht}
\ee
The explicit form of $R_\tht$ is given by
\be
 R_\tht = \begin{pmatrix} R_\tht^{\rm ch} & & & \\ & R_\tht^{\rm ch} & & \\
 & & R_\tht^{\rm nt} & \\ & & & 1 \end{pmatrix}, 
\ee
where 
\be
 R_\tht^{\rm ch} = \begin{pmatrix} c_\tht^2 & s_\tht^2 
 & \sqrt{2}s_\tht c_\tht \\
 s_\tht^2 & c_\tht^2 & -\sqrt{2}s_\tht c_\tht \\
 -\sqrt{2}s_\tht c_\tht & \sqrt{2}s_\tht c_\tht & c_\tht^2-s_\tht^2
 \end{pmatrix}, \;\;\;\;\;
 R_\tht^{\rm nt} = \begin{pmatrix} c_\tht^2 & s_\tht^2 
 & & \sqrt{2}s_\tht c_\tht \\
 s_\tht^2 & c_\tht^2 & & -\sqrt{2}s_\tht c_\tht \\ & & 1 & \\
 -\sqrt{2}s_\tht c_\tht & \sqrt{2}s_\tht c_\tht & & c_\tht^2-s_\tht^2
 \end{pmatrix}. 
\ee

We can decompose $G_{\mu\nu}^{I\bar{J}}(p,y,y')$ into 
the following two parts. 
\be
 G_{\mu\nu}^{I\bar{J}}(p,y,y')=\brkt{\eta_{\mu\nu}-\frac{p_\mu p_\nu}{p^2}}
 G_{\rm T}^{I\bar{J}}(y,y',\abs{p})
 +\frac{p_\mu p_\nu}{p^2}G_{\rm S}^{I\bar{J}}(y,y',\abs{p}),  
\ee 
where $\abs{p}\equiv\sqrt{-p^2}$. 
The first and second terms correspond to the propagators for 
$A_\mu^{(n)}$ and $A_{\rm S}^{(n)}$, respectively. 
Writing $G_{\rm T}^{I\bar{J}}(y,y',\abs{p})$ as 
\be
 G_{\rm T}^{I\bar{J}}(y,y',\abs{p})
 =\vth(y-y')G_{\rm T>}^{I\bar{J}}(y,y',\abs{p})
 +\vth(y'-y)G_{\rm T<}^{I\bar{J}}(y,y',\abs{p}), 
\ee
the solutions to Eq.(\ref{eq_propagator}) satisfying 
Eqs.(\ref{bdcd_propagator:0}) and (\ref{bdcd_propagator:L}) are given 
in the matrix notation for the indices~$(I,\bar{J})$ by 
\bea
 G_{\rm T<}(y,y',\abs{p}) \eql \cM_0(y,\abs{p})\alp_{\rm T<}(y',\abs{p}), 
 \nonumber\\
 R_\theta G_{\rm T>}(y,y',\abs{p}) \eql \cM_L(y,\abs{p})\alp_{\rm T>}(y',\abs{p}),  
 \label{expr:Gp1}
\eea
where 
\be
 \cM_0 \equiv \begin{pmatrix} \cM_0^{\rm ch} & & & \\
 & \cM_0^{\rm ch} & & \\ & & \cM_0^{\rm nt} & \\
 & & & S_0 \end{pmatrix}, \;\;\;
 \cM_L \equiv \begin{pmatrix} \cM_L^{\rm ch} & & & \\
 & \cM_L^{\rm ch} & & \\ & & \cM_L^{\rm nt} & \\
 & & & S_L \end{pmatrix}, 
 \label{def:cM}
\ee
with
\bea
 \cM_0^{\rm ch} \defa \begin{pmatrix} C_0 & & \\ & S_0 & & \\ & & S_0 
 \end{pmatrix}, \;\;\;\;\;
 \cM_0^{\rm nt} \equiv \begin{pmatrix} C_0 & & & \\ 
 & s_\phi^2C_0+c_\phi^2S_0 & s_\phi c_\phi\brkt{C_0-S_0} & \\
 & s_\phi c_\phi\brkt{C_0-S_0} & c_\phi^2C_0+s_\phi^2S_0 & \\
 & & & S_0 \end{pmatrix}, \nonumber\\
 \cM_L^{\rm ch} \defa \begin{pmatrix} C_L & & \\ & C_L & \\ & & S_L 
 \end{pmatrix}, \;\;\;\;\;
 \cM_L^{\rm nt} \equiv \begin{pmatrix} C_L & & & \\ & C_L & & \\
 & & C_L & \\ & & & S_L \end{pmatrix}. 
 \label{def:cMs}
\eea
The unknown matrix functions~$\alp_{\rm T<}(y',\abs{p})$ and 
$\alp_{\rm T>}(y',\abs{p})$ are determined 
by imposing the following matching conditions at $y=y'$. 
The continuity of $G_{\rm T}$ at $y=y'$ leads to the condition
\be
 G_{\rm T<}(y,y,\abs{p}) = G_{\rm T>}(y,y,\abs{p}), 
 \label{continuity}
\ee
and we obtain from Eq.(\ref{eq_propagator}) the condition
\be
 \left\{\der_yG_{\rm T>}(y,y',\abs{p})
 -\der_yG_{\rm T<}(y,y',\abs{p})\right\}_{y'\to y}
 = e^{2\sgm(y)}.  \label{discontinuity}
\ee
Using these conditions, we obtain the 5D propagators as 
\bea
 G_{\rm T<}(y,y',\abs{p}) \eql 
 e^{2\sgm(L)}\cM_0(y,\abs{p})\cW^{-1}(\abs{p})\cM_L(y',\abs{p})R_\theta, 
 \nonumber\\
 G_{\rm T>}(y,y',\abs{p}) \eql \brc{G_{\rm T<}(y',y,\abs{p})}^t, 
 \label{expr:G_T}
\eea
where 
\bea
 \cW(\abs{p}) \defa e^{-2\sgm(y)+2\sgm(L)}
 \brkt{\cM'_L R_\theta\cM_0-\cM_L R_\theta\cM'_0}(y,\abs{p}) \nonumber\\
 \eql \begin{pmatrix} \cW_{\rm ch}(\abs{p}) & & & \\
 & \cW_{\rm ch}(\abs{p}) & & \\ & & \cW_{\rm nt}(\abs{p}) & \\
 & & & \cW^{\hat{4}\hat{4}}(\abs{p}) \end{pmatrix}  \label{def:cW}
\eea
is $y$-independent from the Wronskian relation~(\ref{Wronskian}). 
The explicit forms of the submatrices~$\cW_{\rm ch}$, $\cW_{\rm nt}$ 
and $\cW^{\hat{4}\hat{4}}$ are calculated as 
\bea
 &&\cW_{\rm ch}(m) = -\begin{pmatrix} c_\tht^2C'_0 & s_\tht^2S'_0 & 
 \frac{\sin\thH}{\sqrt{2}}S'_0 \\ s_\tht^2C'_0 & c_\tht^2S'_0 & 
 -\frac{\sin\thH}{\sqrt{2}}S'_0 \\ \frac{me^\sgm\sin\thH}{\sqrt{2}}C_0 & 
 -\frac{me^\sgm\sin\thH}{\sqrt{2}}S_0 & -me^\sgm\cos\thH S_0
 \end{pmatrix}, \nonumber\\
 &&\cW_{\rm nt}(m) \nonumber\\
 &&= -\begin{pmatrix} c_\tht^2C'_0 & 
 s_\tht^2\brkt{s_\phi^2C'_0+c_\phi^2S'_0} & 
 s_\tht^2s_\phi c_\phi\brkt{C'_0-S'_0} & \frac{\sin\thH}{\sqrt{2}}S'_0 \\
 s_\tht^2C'_0 & c_\tht^2\brkt{s_\phi^2C'_0+c_\phi^2S'_0} & 
 c_\tht^2s_\phi c_\phi\brkt{C'_0-S'_0} & -\frac{\sin\thH}{\sqrt{2}}S'_0 \\
 0 & s_\phi c_\phi\brkt{C'_0-S'_0} & c_\phi^2C'_0+s_\phi^2S'_0 & 0 \\
 \frac{me^\sgm\sin\thH}{\sqrt{2}}C_0 & 
 -\frac{me^\sgm\sin\thH}{\sqrt{2}}\brkt{s_\phi^2C_0+c_\phi^2S_0} & 
 -\frac{me^\sgm\sin\thH}{\sqrt{2}}s_\phi c_\phi\brkt{C_0-S_0} & 
 -me^\sgm\cos\thH S_0 \end{pmatrix}, \nonumber\\
 &&\cW^{\hat{4}\hat{4}}(m) = me^\sgm S_0, 
\eea
where the right-hand sides are evaluated at $y=L$. 

The scalar part~$G_{\rm S}(y,y',\abs{p})$ is obtained 
in a similar way, and related to $G_{\rm T}(y,y',\abs{p})$ as 
\be
 G_{\rm S}(y,y',\abs{p}) = G_{\rm T}(y,y',\abs{p}/\sqrt{\xi}). 
\ee

\subsection{Gauge-scalar propagator}
Next we consider the propagators for the gauge-scalar modes. 
The 5D propagator~$iG_{yy}^{I\bar{J}}(y,y',\abs{p})\equiv
\langle 0|TA_y^I(p,y)A_y^{\bar{J}}(-p,y')|0\rangle$ satisfies 
\be
  \brc{\xi\der_y^2 e^{-2\sgm}-p^2}
 G_{yy}^{I\bar{J}}(y,y',\abs{p}) = e^{2\sgm}\dlt^{I\bar{J}}\dlt(y-y'), 
 \label{eq_G:scalar}
\ee
with boundary conditions, 
\bea
 &&G_{yy}^{I\bar{J}} 
 = s_\phi G_{yy}^{3_{\rm R}\bar{J}}+c_\phi G_{yy}^{B\bar{J}}
 = 0,  \;\;\; (I=\pm_{\rm L},3_{\rm L}) \nonumber\\
 &&\der_y\brc{e^{-2\sgm}G_{yy}^{I\bar{J}}} 
 = \der_y\brc{e^{-2\sgm}\brkt{c_\phi G_{yy}^{3_{\rm R}\bar{J}}
 -s_\phi G_{yy}^{B\bar{J}}}} = 0, \;\;\;
 (I=\pm_{\rm R},\hat{\pm},\hat{3},\hat{4})
 \label{bdcd0:scalar}
\eea
at $y=0$, and 
\bea
 &&\brkt{R_\theta}^{IK}G_{yy}^{K\bar{J}} = 0, \;\;\;
 (I=\pm_{\rm L},\pm_{\rm R},3_{\rm L},3_{\rm R},B) \nonumber\\
 &&\brkt{R_\theta}^{IK}\der_yG_{yy}^{K\bar{J}} = 0, \;\;\;
 (I=\hat{\pm},\hat{3},\hat{4}), 
 \label{bdcdL:scalar}
\eea
at $y=L$. 
These can be solved by the same manner as in the previous subsection. 
We find that $G_{yy}(y,y',\abs{p})$ is related to 
$G_{\rm S}(y,y',\abs{p})$ as 
\bea
 G_{yy<}(y,y',\abs{p}) \eql 
 -\frac{1}{p^2}\der_y\der_{y'}G_{\rm S<}(y,y',\abs{p}),  \nonumber\\
 G_{yy>}(y,y',\abs{p}) \eql 
 -\frac{1}{p^2}\der_y\der_{y'}G_{\rm S>}(y,y',\abs{p}).  
 \label{rel:Gyy-Gs}
\eea

\section{Treatment of the forward-scattering singularity} \label{fwd_singularity}
The S-wave amplitude for the elastic scattering: 
$W_L^++W_L^-\to W_L^++W_L^-$ logarithmically diverges 
because of the singularity of the amplitude~$\cA_{WWWW}$ at $\chi=0$. 
Here we show that this divergence is smeared out by taking into account 
the decay width of the W bosons in the final state. 
The instability of the W boson causes an ambiguity 
in the dispersion relation,  
which can be incorporated in the calculation by additional integrals, 
assuming a certain probability dispersion of the ambiguity. 
These additional integrals soften the divergence of the S-wave amplitude, 
as it is an infrared divergence. 

We assume that the W bosons are exactly on-shell in the initial state 
while they can be slightly off-shell in the final state. 
The 4 momenta are parameterized as 
\bea
 p_1 \eql (E/2,0,0,p_W), \nonumber\\
 p_2 \eql (E/2,0,0,-p_W), \nonumber\\
 p_3 \eql (E/2+\dlt E,(p_W+\dlt p)\sin\chi,0,(p_W+\dlt p)\cos\chi), \nonumber\\
 p_4 \eql (E/2-\dlt E,-(p_W+\dlt p)\sin\chi,0,-(p_W+\dlt p)\cos\chi), 
 \label{off-shell_ps}
\eea
where $p_W\equiv\sqrt{E^2/4-m_W^2}$. 
Thus the invariant masses of the final state particles generically 
deviate from the W boson mass~$m_W$, and are parameterized as 
\be
 p_3^2=-(m_W+\dlt m_3)^2, \;\;\;\;\;
 p_4^2=-(m_W+\dlt m_4)^2. 
\ee
The parameters~$\dlt E$ and $\dlt p$ in Eq.(\ref{off-shell_ps}) 
are then expressed in terms of $\dlt m_3$ and $\dlt m_4$ as 
\bea
 \dlt E \eql \frac{m_W}{E}\brkt{\dlt m_3-\dlt m_4}+\cO(\dlt m^2), \nonumber\\
 \dlt p \eql -\frac{m_W}{2p_W}\brkt{\dlt m_3+\dlt m_4}+\cO(\dlt m^2). 
\eea
We assume that the distributions of $\dlt m_3$ and $\dlt m_4$ are given by 
the Gaussian profile, 
\be 
 P(\dlt m) = \frac{1}{\sqrt{2\pi}\Gm_W}\exp\brkt{-\frac{\dlt m^2}{2\Gm_W^2}},  
\ee
where $\Gm_W$ is the decay width of the W boson. 

The S-wave amplitude~$a_0^{00}[WW](E)$ is now expressed as 
\be
 a_0^{00}[WW](E) = \int_{-1}^1\dr x\int_{-\infty}^\infty\dr(\dlt m_3)
 \int_{-\infty}^\infty\dr(\dlt m_4)\;
 P(\dlt m_3)P(\dlt m_4)\frac{f(E,x)}{t+i\ep}, 
 \label{expr:a0}
\ee
where $x\equiv\cos\chi$, $f(E,x)$ is a regular function of $x$,\footnote{
The function~$f(E,x)$ also depends on $\dlt m_3$ and $\dlt m_4$ 
through the 4 momenta~$p_3,p_4$ and the polarization vectors~$\vep_3,\vep_4$. 
However such $\dlt m_{3,4}$-dependences can be neglected in the following 
discussion because they provide only subdominant contributions. 
} 
and 
the Mandelstam variable~$t$ is given by 
\be
 t \equiv (p_1-p_3)^2 
 = -\dlt E^2+\dlt p^2+2p_W(p_W+\dlt p)(1-x). 
\ee
Now we will show the finiteness of the integral in Eq.(\ref{expr:a0}). 
Let us divide the integral region of $x$ as 
\be
 \int_{-1}^1\dr x = \int_{-1}^{x_0}\dr x+\int_{x_0}^1\dr x, 
 \label{divide_int}
\ee
and take $x_0$ as 
\be
 \frac{\abs{\dlt E^2-\dlt p^2}}{2p_W^2} 
 = \frac{8m_W^2\abs{\dlt m_3\dlt m_4}}{E^4} \simlt \frac{16m_W^2\Gm_W^2}{E^4} 
 \ll 1-x_0 \ll 1.  \label{relation}
\ee
Here we have assumed that $\dlt m_{3,4}\simlt \sqrt{2}\Gm_W$.\footnote{ 
Although larger values of $\dlt m_{3,4}$ are possible, their contributions 
to the integral in Eq.(\ref{expr:a0}) are negligible 
due to the tiny probability~$P(\dlt m)\ll 1$. }
Then we can neglect the instability of the W boson 
and replace $P(\dlt m)$ with the delta function~$\dlt(\dlt m)$  
in the first integral in Eq.(\ref{divide_int}).  
For small $\dlt p$, the second integral is estimated as  
\bea
 \int_{x_0}^1\dr x\;\frac{f(E,x)}{t+i\ep} \sma
 -\frac{f(E,1)}{2p_W^2}\ln\frac{\abs{\dlt E^2-\dlt p^2}}{2p_W^2(1-x_0)}. 
 \label{PVInt}
\eea
Here we have neglected the imaginary part of this integral 
coming from the principal value integral when $\dlt E^2-\dlt p^2>0$ 
because it is not enhanced by large logarithm in contrast to the real part. 
Therefore, Eq.(\ref{expr:a0}) is rewritten as 
\be
 a_0^{00}[WW](E) \simeq \int_{-1}^{x_0}\dr x\;\frac{f(E,x)}{2p_W^2(1-x)}
 -\frac{f(E,1)}{2p_W^2}I_{x_0},  \label{expr:a0_2}
\ee
where 
\be
 I_{x_0} \equiv \int\dr(\dlt m_3)d(\dlt m_4)\;P(\dlt m_3)P(\dlt m_4)
 \ln\frac{\abs{\dlt E^2-\dlt p^2}}{2p_W^2(1-x_0)}.  \label{def:I}
\ee
In the following, we will focus on the high energy region~$E^2\gg m_W^2$. 
Then the integral~(\ref{def:I}) is calculated as 
\bea
 I_{x_0} \sma \int\dr(\dlt m_3)d(\dlt m_4)\;P(\dlt m_3)P(\dlt m_4) 
 \ln\frac{8m_W^2\abs{\dlt m_3\dlt m_4}}{E^4(1-x_0)} \nonumber\\
 \eql \int_0^{2\pi}\dr\omg\int_0^\infty\dr r\;
 \frac{r}{2\pi\Gm_W^2}\exp\brkt{-\frac{r^2}{2\Gm_W^2}}
 \ln\frac{4m_W^2r^2\abs{\sin 2\omg}}{E^4(1-x_0)} \nonumber\\
 \eql \frac{1}{2\pi}\int_0^{2\pi}\dr\omg\;
 \brc{-\gm_{\rm E}+\ln\frac{8m_W^2\Gm_W^2\abs{\sin 2\omg}}{E^4(1-x_0)}} \nonumber\\
 \eql -\gm_{\rm E}+\ln\frac{8m_W^2\Gm_W^2}{E^4(1-x_0)}-\ln 2 
 = \ln\frac{4m_W^2\Gm_W^2}{e^{\gm_{\rm E}}E^4(1-x_0)}.  
 \label{expr:I}
\eea
Here we have moved to the polar coordinate~$(\dlt m_3,\dlt m_4)
=(r\cos\omg,r\sin\omg)$ in the second equality, and used the formulae 
\be
 \int_0^\infty\dr y\;\exp(-y)\ln y = -\gm_{\rm E}, \;\;\;\;\;
 \int_0^{2\pi}\dr \omg\;\ln\abs{\sin2\omg} = -2\pi\ln 2. 
\ee
where $\gm_{\rm E}=0.577\cdots$ is the Euler's constant. 

Here we define 
\be
 x_{\rm cut} = 1-\frac{4m_W^2\Gm_W^2}{e^{\gm_{\rm E}}E^4}
 \simeq 1-1.6\times 10^{-3}\frac{m_W^4}{E^4}.  \label{def:x_cut}
\ee
Then Eq.(\ref{expr:a0_2}) with (\ref{expr:I}) is rewritten as 
\bea
 a_0^{00}[WW](E) \sma \int_{-1}^{x_0}\dr x\;\frac{f(E,x)}{2p_W^2(1-x)}
 +\frac{f(E,1)}{2p_W^2}\ln\frac{1-x_0}{1-x_{\rm cut}} \nonumber\\
 \sma \int_{-1}^{x_{\rm cut}}\dr x\;\frac{f(E,x)}{2p_W^2(1-x)}. 
 \label{expr:a0_3}
\eea
We have used $1-x_{\rm cut}\ll 1-x_0\ll 1$ at the last step. 
(See Eq.(\ref{relation}).) 
Note that the $x_0$-dependences are cancelled and the final result is 
independent of $x_0$. 
This is a corollary of the fact that the division of the integral 
region~(\ref{divide_int}) is just an artificial one. 
Eq.(\ref{expr:a0_3}) means that the effect of the instability of the W bosons 
in the final state is translated into the cut-off~$x_{\rm cut}$ 
in the $x$-integral, which regularizes the divergence as expected.  

\ignore{
The $x_0$-dependence of the first term in Eq.(\ref{expr:a0_2}) is negligible 
except for the logarithmic dependence which is cancelled with 
that of the second term. 
This is a corollary of the fact that the division of the integral 
region~(\ref{divide_int}) is just an artificial one. 
The final expression does not depend on $x_0$ although the approximations 
used in the derivation do. 
Thus we can take an arbitrary value of it in Eq.(\ref{expr:a0_2}) 
with (\ref{expr:I}).  
If we set $x_0$ to 
\be
 x_{\rm cut} \equiv 1-\frac{4m_W^2\Gm_W^2}{e^{\gm_{\rm E}}E^4}
 \simeq 1-1.6\times 10^{-3}\frac{m_W^4}{E^4}, 
\ee
the second term in Eq.(\ref{expr:a0_2}) vanishes. 
This means that the effect of the instability of the W bosons 
in the final state is translated into the cut-off~$x_{\rm cut}$ 
in the $x$-integral, which regularizes the divergence as expected.  
}


\end{document}